\documentclass[amsmath,amssymb,pra,superscriptaddress]{revtex4}
\usepackage{graphicx}
\usepackage{dcolumn}
\usepackage{bm}
\usepackage{subfigure}
\usepackage{booktabs}
\usepackage{url}
\usepackage{color}
\newcounter{one}
\def\bfbra#1{\mbox{\boldmath $#1$}^{\mathrm{T}}}
\def\bfket#1{\mbox{\boldmath $#1$}}

\newcommand{\bracket}[1]{\left\langle #1 \right\rangle}

\newcommand{\affA}{Department of Computational Intelligence and Systems Science,
Tokyo Institute of Technology, 4259-G5-22, Nagatsuta-cho, Midori-ku, Yokohama, Kanagawa 226-8502, Japan}

\begin{document}

\title{Limitations in the spectral method for graph partitioning: \\ Detectability threshold and localization of eigenvectors}

\author{Tatsuro Kawamoto}
\affiliation{\affA}
\author{Yoshiyuki Kabashima}
\affiliation{\affA}
\date{\today}

\begin{abstract}
Investigating the performance of different methods is a fundamental problem in graph partitioning.
In this paper, we estimate the so-called detectability threshold for the spectral method with both unnormalized and normalized Laplacians in sparse graphs. The detectability threshold is the critical point at which the result of the spectral method is completely uncorrelated to the planted partition.
We also analyze whether the localization of eigenvectors affects the partitioning performance in the detectable region.
We use the replica method, which is often used in the field of spin-glass theory, and focus on the case of bisection.
We show that the gap between the estimated threshold for the spectral method and the threshold obtained from Bayesian inference is considerable in sparse graphs, even without eigenvector localization.
This gap closes in a dense limit.
\end{abstract}

\maketitle

\section{Introduction} \label{Introduction}
Over recent decades, significant attention has been paid to the clustering problem on graphs or networks \cite{Fortunato201075}.
Although clustering is sometimes considered as simply an optimization problem, e.g., finding the most efficient partitioning in  parallel computing,
it can also be used to find physically meaningful modules, or communities, among which each vertex or edge shares a common attribute.
The latter technique is usually called community detection, and is often formulated as a discrete optimization problem.
Many methods and algorithms have been developed for clustering graphs \cite{Fortunato201075}, and their performance has been investigated both theoretically \cite{Fortunato2007,Kawamoto2015,Reichardt2008,Decelle2011,Decelle2011a,Mossel2014,Nadakuditi2012,VerSteeg2014,Peixoto2013,Newman2013,Radicchi2013,Radicchi2014,Ronhovde2012,Darst02102014} and experimentally \cite{Hu2012,Leskovec2009,Lancichinetti2009,Lancichinetti2011,Good2010,Aldecoa2013,Hric2014,Darst2014}.

One important theoretical problem concerns the so-called \textit{detectability threshold} \cite{Decelle2011,Nadakuditi2012}.
Suppose that we apply a clustering method to a set of random graphs produced by a generative model with a planted block structure, and examine the degree of correlation between the planted partition and the partition obtained by the method.
The standard generative model is called the stochastic block model or planted partition model.
As the connection between modules gets stronger, i.e., as the block structure gets weaker, the partition given by the clustering method will be less correlated with the planted partition.
The detectability threshold is the critical point of a parameter that indicates the strength of the block structure. Below this threshold,  the partition obtained by the method is completely uncorrelated to the planted partition.
This transition is especially problematic in sparse graphs.
For dense graphs, it has been shown that such a transition does not exist and we are always able infer the planted structure in the large size limit \cite{Condon2001,Onsjo2006,Bickel2009}; as we increase the total number of nodes, the average degree also increases in dense graphs, and therefore we obtain more information about the module assignment of each node.
Unfortunately, this is not the case for sparse graphs because the average degree remains constant independent of the graph size.
Needless to say, the detectability threshold in sparse graphs is a significant problem, as many real networks are sparse.
Note that, although we specify the algorithm, i.e., the spectral method, the existence of the detectability threshold itself is not an algorithmic problem, but the theoretical limit of an objective function.

Although there are many levels of graph clustering, e.g., hierarchical clustering and clustering with overlaps, we focus on undirected graphs without hierarchical structures and consider the graph partitioning problem, i.e., partitioning of a graph into non overlapping modules.
In addition, we focus on the case of bisection in the large-size limit.

As mentioned above, the planted partition is given in the analysis of the detectability threshold. In practice, however, not only  do we not  know the model parameters \textit{a priori} (such as the module sizes and the fraction of edges between modules), we are not even sure whether it is appropriate to assume that the graph was generated by the same mechanism as the stochastic block model in the first place.
Therefore, knowing the detectability of a method does not readily contribute to its practical use.
It does, however, provide an important clue for categorizing the detection method: It reveals the similarity of outcomes between methods through the stochastic block model.
Thus, for this purpose, the stochastic block model should be regarded as a model that gives a measure for comparison.
For example, although block structures exist in many senses \cite{Fortunato201075}, applying all the existing methods to the given data obviously entails a huge cost and is also redundant. If we know which methods tend to give a similar partitioning, we could reduce the cost significantly.
Whereas  performance comparison studies are often experimental, it is promising that theoretical analyses can provide deeper insights.

In this paper, we discuss the detectability threshold of the spectral method.
The spectral method can be employed with three major discrete objective functions: ratio-cut (RatioCut), normalized-cuts (Ncut), and modularity.
The spectral method solves the continuous relaxation of these objective functions as an eigenvalue/eigenvector problem of the corresponding matrix. The unnormalized Laplacian $L$ corresponds to RatioCut, the normalized Laplacian $\mathcal{L}$ corresponds to Ncut, and the so-called modularity matrix $B$ corresponds to modularity.
Nadakuditi and Newman \cite{Nadakuditi2012} predicted that the detectability threshold of the spectral method with the modularity function coincides with the threshold given by Bayesian inference \cite{Decelle2011,Decelle2011a}.
However, this assumes that the average degree $\overline{c}$ is sufficiently high.
It was argued in \cite{Zhang2012} that the estimate in \cite{Nadakuditi2012} may not be precise for sparse graphs and that there exists a gap between the spectral method with modularity and Bayesian inference;
indeed, it was confirmed numerically that the spectral method does not detect the planted partition all the way down to the detectability threshold.

It was later discovered that the spectral method with a non-backtracking matrix \cite{Krzakala2013} was a promising means of filling this gap.
This approach provides a formalism that avoids the emergence of localized eigenvectors, known to be a drawback of the spectral method.
A localized eigenvector is one in which the weight of its elements is concentrated on a few characteristic vertices.
Once the eigenvector used for partitioning becomes localized, the information of the block structure will be washed out.
Thus, preventing this effect will enhance our ability to detect the planted structure.

To the best of our knowledge, however, the true detectability threshold for the spectral method in sparse graphs remains unknown. Therefore, we do not know to what extent the gap actually exists.
Furthermore, although the non-backtracking matrix approach improves the detectability by avoiding the eigenvector localization problem, the logical and quantitative connection to detectability seems to be incomplete.
That is, it is not  known whether the gap appears because of the localization or if it exists even when the localization is absent. When both are present, evaluating the relative degree of their effects is an important problem.
Our results show that, in sparse graphs, a considerable gap exists even when localization is absent and that the effect of  localization may be significant when the degree fluctuates considerably.

Using the so-called \textit{replica method}, which is often used in spin-glass theory, we derive estimates for the detectability threshold of the spectral method with both unnormalized and normalized Laplacians.
Note that, as pointed out in \cite{Newman2013}, the spectral method with the normalized Laplacian $\mathcal{L}$ and the modularity matrix $B$ are equivalent for the bisection problem, as long as the continuous relaxation gives a good estimate of the original discrete problem.
We compare our analytical estimates with the results of numerical experiments for the two-block random graphs with uniform, bimodal, and Poisson degree distributions.
Although our estimates contain some approximations, they agree quite well with the results of the numerical experiment, as long as localization does not occur.
For the analysis of localized eigenvectors, we show that our estimate is fairly accurate for graphs with bimodal degree distributions.

The rest of this paper is organized as follows.
We first introduce a more precise definition of the stochastic block model (Sec.~\ref{StochasticBlockModel}) and the spectral method in graph partitioning (Sec.~\ref{SpectralMethods}).
Then we derive an estimate of the detectability threshold in two-block random regular graphs in Sec.~\ref{DetectabilityRegularRandom}.
Note that there is no distinction between unnormalized and normalized Laplacians in this case.
We analyze the effect of degree fluctuation for the unnormalized Laplacian $L$ in Secs.~\ref{DetectabilityRatioCut} and \ref{LocalizationRatioCut}.
In Sec.~\ref{DetectabilityRatioCut} we present a formal solution for estimating the detectability threshold and analyze the case of graphs with bimodal degree distributions, and in Sec.~\ref{LocalizationRatioCut} we  estimate the localized eigenvector and its eigenvalue for a graph with a bimodal degree distribution.
A similar analysis is done for the normalized Laplacian $\mathcal{L}$ in Secs.~\ref{DetectabilityNcut} and \ref{LocalizationNcut}.
For the normalized Laplacian $\mathcal{L}$ of graphs with an arbitrary degree distribution, the resulting estimate of the detectability threshold resembles that of random regular graphs.
Finally, we discuss the case of stochastic block models, i.e., two-block random graphs with Poisson degree distributions, with the normalized Laplacian $\mathcal{L}$ in Sec.~\ref{DetectabilitySBM}.
We summarize our results in Sec.~\ref{Summary}.

\section{Stochastic block model and the detectability threshold} \label{StochasticBlockModel}
The stochastic block model \cite{holland1983stochastic} is a generative model of random graphs with a block structure, and is commonly used for analyzing the performance of clustering methods.
While many variants have been proposed \cite{Airoldi2008,Clauset2008,Karrer2011,Ball2011,Peixoto2012,Larremore2014},
the model is fundamentally a generalization of the Erd{\H{o}}s-R\'{e}nyi random graph.
In the stochastic block model, the number of modules $q$, size of each module, and probability $p_{rs}$ that vertices in modules $r$ and $s$ are connected are specified as inputs.
With these parameters, the graphs are constructed as follows.
Each vertex has a preassigned module index $\sigma_{i} = r \, (r \in \{1, \dots, q\})$ to which the vertex belongs. Based on this block structure, edges are generated between pairs of vertices at random, i.e., vertices $i \in r$ and $j \in s$  are connected with probability $p_{rs}$.

In the case of sparse graphs of $N$ vertices, we set $p_{rs} = c_{rs}/N$, where $c_{rs}$ remains constant in the limit $N \rightarrow \infty$.
To construct an assortative block structure, we typically choose $p_{rr} = c_{\mathrm{in}}/N$ and $p_{rs} = c_{\mathrm{out}}/N$ for $r \ne s$, where $c_{\mathrm{in}}$ and $c_{\mathrm{out}}$ are constants that satisfy $c_{\mathrm{in}} > c_{\mathrm{out}}$.
In the case of two modules of equal size, we have the average degree $\overline{c} = (c_{\mathrm{in}}+c_{\mathrm{out}})/2$.
Note that, in this case, unlike the dense case, the fluctuation of the degree of each vertex does not vanish in the limit $N \rightarrow \infty$.

According to \cite{Decelle2011,Decelle2011a}, the Bayesian inference method has a detectability threshold at
\begin{align}
c_{\mathrm{in}} - c_{\mathrm{out}} = 2 \sqrt{\overline{c}}.
\end{align}
That is, even when the generative model has the assortative property $c_{\mathrm{in}} > c_{\mathrm{out}}$, it is impossible to retrieve that information with any detection algorithm, unless the difference $c_{\mathrm{in}} - c_{\mathrm{out}}$ is greater than $2 \sqrt{\overline{c}}$
(for other detectability analyses with sufficiently large average degree, see \cite{Radicchi2013,Radicchi2014,Zhang2014,ChenHero2014}).

In the present paper, we parametrize the stochastic block model differently. As mentioned above, we restrict ourselves to the two-block model.
Instead of setting $c_{\mathrm{in}}$ and $c_{\mathrm{out}}$ as the model parameters, we set the average degree $\overline{c}$ and the average number of edges $\gamma$ from one module to the other, or, in other words, the total number of edges $\gamma N$ between two modules.
The parameters $\gamma$ controls the strength of block structure; the larger is the value of $\gamma$, the weaker is the block structure. 
It is related to $c_{\mathrm{out}}$ and $c_{\mathrm{in}}$ by
\begin{align}
\gamma N &= \left( \frac{N}{2} \right)^{2} p_{\mathrm{out}}
= \frac{N}{4} c_{\mathrm{out}}, \label{Kout} \\
\frac{\overline{c}N}{2} - \gamma N &= 
2
\begin{pmatrix}
\frac{N}{2} \\
2
\end{pmatrix}
= \frac{N}{4} c_{\mathrm{in}}, \label{Kin}
\end{align}
in the limit $N\rightarrow\infty$. 
Although the total number of edges between modules fluctuates in the standard formulation when the graphs are finite, we let every realization have exactly $\gamma N$ and $\overline{c}N/2 - \gamma N$ for the number of edges between modules and the total number of edges within modules, respectively. 
This is called the ``microcanonical'' formulation of the stochastic block model \cite{Peixoto2012}.

\section{Spectral method in graph bisection} \label{SpectralMethods}
Graph partitioning is often formulated as a discrete optimization problem for some objective function that is computationally difficult.
The spectral method constitutes a continuous relaxation of the original problem using eigenvectors of a proper matrix.
The unnormalized Laplacian $L$, which is used for RatioCut, and the normalized Laplacian $\mathcal{L}$, which is used for Ncut, are defined as
\begin{align}
&L = D - A, \label{unnormalizedLaplacian} \\
&\mathcal{L} = D^{-1/2} L D^{-1/2}. \label{unnormalizedLaplacian}
\end{align}
The matrix $A$ is the adjacency matrix, i.e., $A_{ij}=1$ if vertices $i$ and $j$ are connected and $A_{ij}=0$ otherwise. The matrix $D$ is a diagonal matrix with degree $c_{i}$ of vertex $i$ on the diagonal element, i.e., $D_{ij} = c_{i} \delta_{ij}$.

Although the details differ depending on the objective function, the basic procedure of spectral bisection is the same and quite simple \cite{Luxburg2007,Newman2013}.
We denote  the total degree of the graph as $K$.
For a graph with a set of vertices $V$ partitioned into $V_{1}$ and $V_{2}$,
the objective function of RatioCut is defined as
 \begin{align}
f_{\mathrm{RatioCut}}(V_{1}, V_{2})
&= \frac{E(V_{1}, V_{2})}{N_{1} N_{2}}, \label{RatioCut}
\end{align}
where the cut size $E(V_{1}, V_{2})$ is the number of edges between modules $V_{1}$ and $V_{2}$,
and we denote by $N_{1}$ and $N_{2}$ the number of vertices in each module.
Similarly, the objective function of Ncut is defined as
\begin{align}
f_{\mathrm{Ncut}}(V_{1}, V_{2})
&= \frac{E(V_{1}, V_{2})}{K_{1} K_{2}}, \label{Ncut}
\end{align}
where $K_{1}$ and $K_{2}$ are the total degrees of each modules.
Using the unnormalized Laplacian $L$ and the normalized Laplacian $\mathcal{L}$, minimizing RatioCut and Ncut are equivalent to
\begin{align}
\min_{\bfket{x}} \bfbra{x} L \bfket{x}, \hspace{10pt}
\text{subject to} \hspace{10pt}
x_{i} =
\begin{cases}
\sqrt{ N_{2} / N_{1} } & i \in V_{1}, \\
- \sqrt{ N_{1} / N_{2} } & i \in V_{2},
\end{cases} \label{LaplacianRatioCut}
\end{align}
and
\begin{align}
\min_{\bfket{x}} \bfbra{x} \mathcal{L} \bfket{x}, \hspace{10pt}
\text{subject to} \hspace{10pt}
x_{i} =
\begin{cases}
\sqrt{ K_{2} / K_{1} } & i \in V_{1}, \\
- \sqrt{ K_{1} / K_{2} } & i \in V_{2}.
\end{cases} \label{LaplacianNcut}
\end{align}
Finally, allowing $x_{i}$ to take an arbitrary real number, we obtain the relaxed versions of the above discrete optimization problems.
For RatioCut,
\begin{align}
\min_{\bfket{x} \in \mathbb{R}^{N}} \bfbra{x} L \bfket{x} \hspace{10pt}
\text{subject to} \hspace{10pt}
\bfket{x} \perp \bfket{1}, \,\, \bfbra{x}\bfket{x} = N, \label{relaxedRatioCut}
\end{align}
and for Ncut,
\begin{align}
\min_{\bfket{x} \in \mathbb{R}^{N}} \bfbra{x} \mathcal{L} \bfket{x} \hspace{10pt}
\text{subject to} \hspace{10pt}
\bfket{x} \perp D^{1/2} \bfket{1}, \,\, \bfbra{x}\bfket{x} = K. \label{relaxedNcut}
\end{align}
The smallest values of $\bfbra{x} L \bfket{x}$ and $\bfbra{x} \mathcal{L} \bfket{x}$ are achieved when the $\bfket{x}$ are the eigenvectors corresponding to the smallest eigenvalue of $L$ and $\mathcal{L}$, respectively.
Note, however, that $L$ is positive semi-definite and $\bfket{1}$ is the eigenvector of $L$ corresponding to the zero eigenvalue (i.e., the smallest eigenvalue).
Because of the constraint that $\bfket{x}$ must be perpendicular to $\bfket{1}$, the smallest value of $\bfbra{x} L \bfket{x}$ in Eq.~(\ref{relaxedRatioCut}) is achieved when we select the eigenvector corresponding to the second-smallest eigenvalue of $L$.
Similarly, $\mathcal{L}$ is also positive semi-definite, and $D^{1/2} \bfket{1}$ is the eigenvector corresponding to the zero eigenvalue.
Hence, the eigenvector corresponding to the second-smallest eigenvalue of $\mathcal{L}$ gives the smallest value of $\bfbra{x} \mathcal{L} \bfket{x}$ in Eq.~(\ref{relaxedNcut}).
As the sign of $x_{i}$ indicates which module vertex $i$ belongs to in Eqs.~(\ref{LaplacianRatioCut}) and (\ref{LaplacianNcut}), we retrieve information about the optimal partition from the solutions of the relaxed problems by referring to the sign of each element in the eigenvector, i.e., vertices with the same sign belong to the same module. It is known that this prescription works well when the module sizes are not very different.
Of course, it is not obvious whether the optimal partition of the relaxed problem coincides with that of the unrelaxed problem.
However, this is beyond the scope of the present paper, and we concentrate on the relaxed problem, i.e., the spectral method.

In the following sections, we analyze how the optimal partitions in the spectral method are correlated to the planted partitions in various random graphs.

\section{Detectability threshold in random regular graphs}\label{DetectabilityRegularRandom}

\begin{figure}[t]
\centering
\includegraphics[width=0.4\columnwidth]{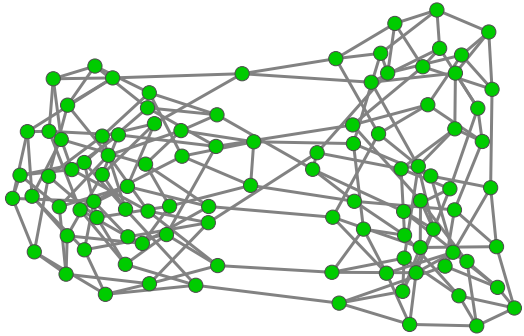}
\caption{(Color online) A realization of two-block $4$-random regular graphs.
}
\label{RegularRandomFigure}
\end{figure}

In this section, we analyze the detectability of the $c$-random regular graph with a two-block structure.
A realization of such graphs is shown in Fig.~\ref{RegularRandomFigure}.
As $c$ is a constant and does not increase as a function of the graph size $N$, this is regarded as a sparse random graph that has the property of dense graphs that the degree fluctuation is negligible.
This is worth investigating, because the results we show in Sec.~\ref{DetectabilityNcut} for the normalized Laplacian $\mathcal{L}$ are analogous to those we obtain for random regular graphs in this section.
Although we analyze the unnormalized Laplacian $L$ here, as the degree is the same for every vertex, there is no distinction  between the unnormalized and normalized Laplacians.

We calculate the average of the second-smallest eigenvalue of the unnormalized Laplacian $\left[ \lambda_{2} \right]_{L}$, where $\left[ \dots \right]_{L}$ represents the average with respect to $L$ of each realization of a random graph.
As we increase the fraction $\gamma$ of edges between modules, the value of $\left[ \lambda_{2} \right]_{L}$ increases until it reaches the edge of the spectral band, above which $\left[ \lambda_{2} \right]_{L}$ becomes constant irrespective of $\gamma$.
As we show in the following, by calculating $\left[ \lambda_{2} \right]_{L}$, we can obtain the distribution of the elements in the corresponding eigenvector.

The basic methodology here runs parallel to that in \cite{Kabashima2012}.
To calculate the second-smallest eigenvalue, we first introduce the Hamiltonian $H(\bm{x} | L)$, partition function $Z(\beta | L)$, and free energy $f(\beta | L)$, 
\begin{align}
&H(\bm{x} | L) = \frac{1}{2} \bm{x}^{\mathrm{T}} L \bm{x}, \label{RatioCutHamiltonian}\\
&Z(\beta | L) = \int d \bm{x} \, \mathrm{e}^{-\beta H(\bm{x} | L)} \delta(|\bm{x}|^{2} - N) \delta(\bm{1}^{\mathrm{T}} \bm{x}), \label{RatioCutPartitionFunction}\\
&f(\beta | L) = -\frac{1}{N\beta} \ln Z(\beta | L), \label{RatioCutFreeEnergy}
\end{align}
where $\bm{x}$ is an $N$-dimensional vector, $\bm{1}$ is the vector in which each element is $1$, and $\mathrm{T}$ denotes the transpose.
The Hamiltonian corresponds to the objective function to be minimized in Eq.~(\ref{relaxedRatioCut}); the factor $1/2$ in Eq.~(\ref{RatioCutHamiltonian}) is purely conventional.
The $\delta$ functions in Eq.~(\ref{RatioCutPartitionFunction}) impose the constraints in Eq.~(\ref{relaxedRatioCut}).
The crucial aspect of this formulation is that, in the limit $\beta \rightarrow \infty$, in conjunction with the operation of $\delta(\bfbra{1} \bfket{x})$, the contribution from the second-smallest eigenvalue is dominant in the integral in Eq.~(\ref{RatioCutPartitionFunction}). Thus, Eq.~(\ref{RatioCutPartitionFunction}) actually evaluates $\exp\left[ -N \beta \lambda_{2} / 2 \right]$.
Hence, the second-smallest eigenvalue $\lambda_{2}$ is given by
\begin{align}
\lambda_{2} = 2 \lim_{\beta \rightarrow \infty} f(\beta | L).
\end{align}
We then take the average over all realizations of random graphs.
However, the direct calculation of this average is not tractable. Therefore, we recast $\left[\lambda_{2}\right]_{L}$ as
\begin{align}
\left[\lambda_{2}\right]_{L}
&= -2 \lim_{\beta \rightarrow \infty} \frac{1}{N\beta} \left[\ln Z(\beta | L)\right]_{L}\nonumber\\
&= -2 \lim_{\beta \rightarrow \infty} \lim_{n\rightarrow 0} \frac{1}{N\beta} \frac{\partial}{\partial n} \ln \left[Z^{n}(\beta | L)\right]_{L}. \label{ReplicaTrick}
\end{align}
The assessment of $[Z^{n}(\beta|L)]_{L}$ is also difficult for a general real number $n$. However, when $n$ is a positive integer, $[Z^{n}(\beta|L)]_{L}$ can be expressed as a high-dimensional integral with respect to $n$ replicated variables $\bm{x}^{1}, \bm{x}^{2}, \dots, \bm{x}^{n}$, which can be analytically evaluated by the saddle-point method as $N$ tends to infinity. In addition, the resulting expression of $N^{-1} \ln [Z^{n}(\beta|L)]_{L}$ is shown to be a function of $n$ that can be extended to real values of $n$ under a certain ansatz concerning the permutation symmetry among the replica indices $a=1, 2, \dots, n$. Therefore, we employ such an expression to calculate the right-hand side of Eq.~(\ref{ReplicaTrick}). This procedure is often termed the replica method. Although the mathematical validity of the replica method has not yet been proved, we see that our assessment based on the simplest permutation symmetry for the replica indices offers a fairly accurate prediction of the experimental results.

For $n \in \mathbb{N}$, we can write $\left[Z^{n}(\beta | L)\right]_{L}$ as
\begin{align}
[Z^{n}(\beta | L)]_{L}
= \int \left(\prod_{a=1}^{n} d \bfket{x}_{a} \delta(|\bfket{x}_{a}|^{2} - N) \delta(\bfbra{1} \bfket{x}_{a}) \right) \left[ \exp\left(-\frac{\beta}{2} \sum_{a} \bfbra{x}_{a} L \bfket{x}_{a} \right) \right]_{L}, \label{Z^n}
\end{align}
and the exponential factor is given by
\begin{align}
&\exp\left(-\frac{\beta}{2} \sum_{a} \bm{x}_{a}^{\mathrm{T}} L \bm{x}_{a} \right) \nonumber\\
&\hspace{20pt}= \exp\left[ -\frac{\beta}{4} \sum_{a=1}^{n} \left( \sum_{ij \in V_{1}} u_{ij}(x_{ia} - x_{ja})^{2} + \sum_{ij \in V_{2}} u_{ij}(x_{ia} - x_{ja})^{2} + 2\sum_{i \in V_{1}}\sum_{j \in V_{2}} w_{ij}(x_{ia} - x_{ja})^{2} \right) \right], \label{BoltzmannFactor1}
\end{align}
where $u_{ij} = \{ A_{ij} | i \in r, j \in r \}$ and $w_{ij} = \{ A_{ij} | i \in V_{1}, j \in V_{2} \}$.
For the average over the random graphs, we assume that each realization occurs with equal probability.
The condition of being a regular graph requires
\begin{align}
&\sum_{l \in V_{1}} u_{il} + \sum_{k \in V_{2}} w_{ik} = c \hspace{20pt} \text{(for $i \in V_{1}$)}, \\
&\sum_{l \in V_{2}} u_{jl} + \sum_{k \in V_{1}} w_{jk} = c \hspace{20pt} \text{(for $j \in V_{2}$)},
\end{align}
and the number of edges between modules is
\begin{align}
\sum_{i \in V_{1}} \sum_{k \in V_{2}} w_{ik} = \gamma N.
\end{align}
Therefore, we have
\begin{align}
&\left[ \exp\left(-\frac{\beta}{2} \sum_{a} \bm{x}_{a}^{\mathrm{T}} L \bm{x}_{a} \right) \right]_{L} \nonumber\\
&\hspace{10pt}= \frac{1}{\mathcal{N}_{G}} \sum_{ \{u_{ij}\} \{w_{ij}\} }
\Biggl\{
\prod_{i \in V_{1}} \delta\left( \sum_{l \in V_{1}} u_{il} + \sum_{k \in V_{2}} w_{ik} - c \right)
\prod_{j \in V_{2}} \delta\left( \sum_{l \in V_{2}} u_{jl} + \sum_{k \in V_{1}} w_{jk} - c \right)
\delta\left( \sum_{i \in V_{1}} \sum_{k \in V_{2}} w_{ik} - \gamma N \right) \nonumber\\
&\hspace{20pt} \times
\exp\left[ -\frac{\beta}{4} \sum_{a=1}^{n} \left( \sum_{ij \in V_{1}} u_{ij}(x_{ia} - x_{ja})^{2} + \sum_{ij \in V_{2}} u_{ij}(x_{ia} - x_{ja})^{2} + 2\sum_{i \in V_{1}}\sum_{j \in V_{2}} w_{ij}(x_{ia} - x_{ja})^{2} \right) \right]
\Biggr\}, \label{BoltzmannFactor2}
\end{align}
where $\mathcal{N}_{G}$ is the number of random regular graphs with two modules (see Appendix~\ref{GraphCountSingleDegree} for the count of $\mathcal{N}_{G}$ in random regular graphs).
With the technique in Appendix~\ref{GraphCountSingleDegree}, Eq.~(\ref{BoltzmannFactor2}) can be written as
\begin{align}
\left[ \exp\left(-\frac{\beta}{2} \sum_{a} \bm{x}_{a}^{\mathrm{T}} L \bm{x}_{a} \right) \right]_{L}
&= \frac{1}{\mathcal{N}_{G}}
\oint \prod_{i \in V_{1}} \frac{dz_{i}}{2\pi} z_{i}^{-(1+c)}
\oint \prod_{j \in V_{2}} \frac{dz_{j}}{2\pi} z_{j}^{-(1+c)}
\int \frac{d\eta}{2\pi} \mathrm{e}^{\eta \gamma N} \nonumber\\
&\hspace{20pt} \times
\prod_{i<j \in V_{1}} \sum_{u_{ij} = \{0,1\}} \left( z_{i}z_{j} \exp\left[ -\frac{\beta}{2} \sum_{a} (x_{ia} - x_{ja})^{2} \right] \right)^{u_{ij}} \nonumber\\
&\hspace{20pt} \times \prod_{i<j \in V_{2}} \sum_{u_{ij} = \{0,1\}} \left( z_{i}z_{j} \exp\left[ -\frac{\beta}{2} \sum_{a} (x_{ia} - x_{ja})^{2} \right] \right)^{u_{ij}} \nonumber\\
&\hspace{20pt} \times \prod_{i \in V_{1}}\prod_{j \in V_{2}} \sum_{w_{ij} = \{0,1\}} \left( z_{i}z_{j} \exp\left[ -\eta -\frac{\beta}{2} \sum_{a} (x_{ia} - x_{ja})^{2} \right] \right)^{w_{ij}}. \label{BoltzmannFactor3}
\end{align}
When $N \gg 1$, an element in Eq.~(\ref{BoltzmannFactor3}) can be approximated as
\begin{align}
&\prod_{i<j \in V_{r}} \sum_{u_{ij} = \{0,1\}} \left( z_{i}z_{j} \exp\left[ -\frac{\beta}{2} \sum_{a} (x_{ia} - x_{ja})^{2} \right] \right)^{u_{ij}} \nonumber\\
&\hspace{20pt} \approx \exp \left[ \sum_{i<j \in V_{r}} z_{i}z_{j} \exp\left( -\frac{\beta}{2} \sum_{a} (x_{ia} - x_{ja})^{2} \right) \right] \nonumber\\
&\hspace{20pt} \approx \exp \left[ \frac{(p_{1}N)^{2}}{2} \int \prod_{a=1}^{n} d\mu^{(r)}_{a}d\nu^{(r)}_{a}
 \mathcal{Q}_{r}(\bm{\mu}^{(r)}) \mathcal{Q}_{r}(\bm{\nu}^{(r)})
 \exp\left( -\frac{\beta}{2} \sum_{a} (\mu^{(r)}_{a} - \nu^{(r)}_{a})^{2} \right)
 \right],
\end{align}
where we have introduced the order parameter functions
\begin{align}
\mathcal{Q}_{r}(\bm{\mu}^{(r)}) = \frac{1}{p_{r}N} \sum_{i \in V_{r}} z_{i} \prod_{a} \delta(x_{ia} - \mu^{(r)}_{a}),
\end{align}
for $r =1, 2$.
Then, inserting the identity
\begin{align}
1 &= \int d\mathcal{Q}_{r}(\bm{\mu}^{(r)}) \delta \left( \frac{1}{p_{r} N} \sum_{i \in V_{r}}z_{i}\prod_{a}\delta(x_{ia} - \mu^{(r)}_{a}) - \mathcal{Q}_{r}(\bm{\mu}^{(r)}) \right) \nonumber\\
&= p_{r}N \int \frac{d\mathcal{Q}_{r}(\bm{\mu}^{(r)}) d\hat{\mathcal{Q}}_{r}(\bm{\mu}^{(r)})}{2\pi}
\exp \left[ \hat{\mathcal{Q}}_{r}(\bm{\mu}^{(r)}) \left( \sum_{i \in V_{r}}z_{i}\prod_{a}\delta(x_{ia} - \mu^{(r)}_{a}) - p_{r} N \mathcal{Q}_{r}(\bm{\mu}^{(r)}) \right) \right]
\end{align}
for each $\bm{\mu}^{(r)}$ and replacing the $\delta$ functions as
\begin{align}
\delta\left( \sum_{i=1}^{N}x^{2}_{ia} - N \right)
&= \int \frac{\beta d\phi_{a}}{4\pi} \, \mathrm{e}^{-\frac{\beta}{2}\phi_{a}(\sum_{i} x^{2}_{ia} - N)}, \label{NormalizationConstraint}\\
\delta\left( \sum_{i=1}^{N}x_{ia} \right)
&= \int \frac{\beta d\psi_{a}}{4\pi} \, \mathrm{e}^{-\frac{\beta}{2}\psi_{a}(\sum_{i} x_{ia})}, \label{OrthogonalityConstraint}
\end{align}
we can recast Eq.~(\ref{Z^n}) as
\begin{align}
[Z^{n}(\beta | L)]_{L}
&= p_{1} p_{2} N^{2} \int \prod_{r=1,2} \frac{d\mathcal{Q}_{r}(\bm{\mu}^{(r)})d\hat{\mathcal{Q}}_{r}(\bm{\mu}^{(r)})}{2\pi}
\int \prod^{n}_{a} \frac{\beta d\phi_{a}}{4\pi} \frac{\beta d\psi_{a}}{4\pi} \int \frac{d\eta}{2\pi} \nonumber\\
&\hspace{20pt}\times 
\exp \Biggl[ N^{2} K_{\mathrm{I}}(\mathcal{Q}_{r}, \hat{\mathcal{Q}}_{r}) + N \biggl( \frac{\beta}{2} \sum_{a} \phi_{a}
- \sum_{r = 1,2} K_{\mathrm{II} r}(\mathcal{Q}_{r}, \hat{\mathcal{Q}}_{r}) + \frac{1}{N} \sum_{r = 1,2} \ln K_{\mathrm{III},r}(\hat{\mathcal{Q}}_{r}, \{\phi_{a}\}, \{\psi_{a}\}) \nonumber\\
&\hspace{50pt} + \eta \gamma - \frac{1}{N} \ln \mathcal{N}_{G} - \ln c ! \biggr) \Biggr],
\end{align}
where
\begin{align}
&K_{\mathrm{I}}(\mathcal{Q}_{r}, \hat{\mathcal{Q}}_{r}) = \sum_{r,s = 1,2} \frac{p_{r}p_{s}}{2} \int d\bm{\mu}^{(r)} d\bm{\nu}^{(s)} \, \mathcal{Q}_{r}(\bm{\mu}^{(r)}) \mathcal{Q}_{s}(\bm{\nu}^{(s)}) \mathrm{e}^{-(1-\delta_{rs})\eta -\frac{\beta}{2}\sum_{a}(\mu^{(r)}_{a}-\nu^{(s)}_{a})^{2} }, \label{KI} \\
&K_{\mathrm{II} r}(\mathcal{Q}_{r}, \hat{\mathcal{Q}}_{r}) = p_{r} \int d\bm{\mu}^{(r)} \, \hat{\mathcal{Q}}_{r}(\bm{\mu}^{(r)}) \mathcal{Q}_{r}(\bm{\mu}^{(r)}), \label{KII}\\
&K_{\mathrm{III} r}(\hat{\mathcal{Q}}_{r}, \{\phi_{a}\}, \{\psi_{a}\})
= \int \prod_{i \in V_{r}} \prod_{a=1}^{n} dx_{ia} \, \prod_{i \in V_{r}} \left( \hat{\mathcal{Q}}^{c}_{r}(\bm{x}_{i})
\exp \left[ -\frac{\beta}{2} \sum_{a} \left( \phi_{a} x^{2}_{ia} + \psi_{a} x_{ia} \right) \right] \right). \label{KIII}
\end{align}

Now we evaluate $\lim_{N\rightarrow\infty} (\ln [Z^{n}(\beta | L)]_{L})/N$ with the saddle-point method and calculate the second-smallest eigenvalue according to Eq.~(\ref{ReplicaTrick}).
For this, we assume that the functional forms of $\mathcal{Q}_{r}(\mu)$ and $\hat{\mathcal{Q}}_{r}(\mu)$ are invariant under any permutations of replica indices $a\in \{1,2,\dots,n\}$, which is often termed the replica-symmetric ansatz. Further, the Gaussian nature of the current problem allows us to assume that $\mathcal{Q}_{r}(\mu)$ and $\hat{\mathcal{Q}}_{r}(\mu)$ are mixtures of Gaussian functions; this originates from the fact that the effective Hamiltonian yielded by appropriate exponentiations of the $\delta$ functions is composed of quadratic forms. These restrict the functional forms of $\mathcal{Q}_{r}(\mu)$ and $\hat{\mathcal{Q}}_{r}(\mu)$ as
\begin{align}
&\mathcal{Q}_{r}(\bm{\mu}) = T_{r} \int dA dH \, q_{r}(A,H) \left( \frac{\beta A}{2\pi} \right)^{\frac{n}{2}}
\exp\left[ -\frac{\beta A}{2} \sum_{a=1}^{n} \left( \mu_{a} - \frac{H}{A} \right)^{2} \right], \label{assumedQ}\\
&\hat{\mathcal{Q}}_{r}(\bm{\mu}) = \hat{T}_{r} \int d\hat{A} d\hat{H} \, \hat{q}_{r}(\hat{A},\hat{H})
\exp\left[ \frac{\beta}{2} \sum_{a=1}^{n} \left( \hat{A} \mu^{2}_{a} + 2 \hat{H} \mu_{a} \right) \right], \label{assumedhatQ}
\end{align}
i.e., some superpositions of Gaussian functions with weights $q_{r}(A,H)$ and $\hat{q}_{r}(\hat{A},\hat{H})$, where $A$ and $\hat{A}$ denote the variances and $H$ and $\hat{H}$ denote the means of each Gaussian distribution, respectively.
When $n=0$, the order parameters $\mathcal{Q}_{r}(\bm{\mu})$ and $\hat{\mathcal{Q}}_{r}(\bm{\mu})$ coincide with Eqs.~(\ref{Appendix-OrderParameter1}) and (\ref{Appendix-OrderParameter2}) in Appendix~\ref{GraphCountSingleDegree} and thus the normalization factors $T_{r}$ and $\hat{T}_{r}$ are equal to Eqs.~(\ref{T1}) and (\ref{T2}).
With Eqs.~(\ref{assumedQ}) and (\ref{assumedhatQ}),
(\ref{KI})--(\ref{KIII}) become functions of $n$ that are extendable to real values of $n$. Inserting these expressions into the identity $N^{-1} \left[\ln Z(\beta | L)\right]_{L} = \lim_{n \to 0} (\partial/\partial n) N^{-1} \ln \left[Z^{n}(\beta | L)\right]_{L}$, we have
\begin{align}
\left[\lambda_{2}\right]_{L}
&= - \mathop{\mathrm{extr}}_{q_{r}, \hat{q}_{r}, \phi, \psi} \Biggl\{ \int dA dH \int dA^{\prime} dH^{\prime} \, \Xi(A, H, A^{\prime}, H^{\prime}) \nonumber\\
&\hspace{10pt} \times \frac{cp_{1}p_{2}}{2} \biggl(
\left(\frac{p_{1}}{p_{2}} + \Gamma\right) q_{1}(A, H) q_{1}(A^{\prime}, H^{\prime}) +
\left(\frac{p_{2}}{p_{1}} + \Gamma\right) q_{2}(A, H) q_{2}(A^{\prime}, H^{\prime}) +
2 \left(1 - \Gamma\right) q_{1}(A, H) q_{2}(A^{\prime}, H^{\prime})
\biggr)
\nonumber\\
&\hspace{10pt} +\phi - c \sum_{r = 1,2} p_{r} \int dA dH \int d\hat{A} d\hat{H} \,
q_{r}(A, H) \hat{q}_{r}(\hat{A}, \hat{H})
\left(
\frac{(H + \hat{H})^{2}}{A - \hat{A}} - \frac{H^{2}}{A}
\right) \nonumber\\
&\hspace{20pt} + \sum_{r=1,2} p_{r} \int \prod_{g=1}^{c} \left( d\hat{A}_{g}d\hat{H}_{g} \hat{q}_{r}(\hat{A}_{g}, \hat{H}_{g}) \right)
\frac{\left(\psi/2 - \sum_{g}\hat{H}_{g}\right)^{2}}{\phi - \sum_{g}\hat{A}_{g}}
\Biggr\}, \label{2ndEigenValue}
\end{align}
where we set
\begin{align}
&\Gamma = 1 - \frac{\gamma}{c p_{1} p_{2}}, \label{GammaRegular} \\
&\Xi(A, H, A^{\prime}, H^{\prime}) = \frac{(1+A^{\prime})H^{2}+(1+A)H^{\prime 2}+2HH^{\prime}}{(1+A)(1+A^{\prime})-1}
- \frac{H^{2}}{A} - \frac{H^{\prime 2}}{A^{\prime}}.
\end{align}
In the above calculation, we have assumed the replica symmetry for $\phi_{a}$ and $\psi_{a}$, i.e., $\phi_{a}=\phi$ and $\psi_{a} = \psi$, respectively, for $a=1,2,\dots,n$.
The saddle-point conditions yield the following set of integral equations:
\begin{align}
\hat{q}_{1}(\hat{A}, \hat{H})
= \int dA^{\prime} dH^{\prime} \, \left[ \left( 1 + \frac{p_{2}}{p_{1}}\Gamma \right) p_{1} q_{1}(A^{\prime}, H^{\prime}) + \left( 1 - \Gamma \right) p_{2} q_{2}(A^{\prime}, H^{\prime}) \right]
\delta\left( \hat{A} + \frac{A^{\prime}}{1 + A^{\prime}} \right)
\delta\left( \hat{H} - \frac{H^{\prime}}{1 + A^{\prime}} \right), \label{cavity1-Regular}\\
\hat{q}_{2}(\hat{A}, \hat{H})
= \int dA^{\prime} dH^{\prime} \, \left[ \left( 1+\frac{p_{1}}{p_{2}}\Gamma \right) p_{2} q_{2}(A^{\prime}, H^{\prime}) + \left( 1-\Gamma \right) p_{1} q_{1}(A^{\prime}, H^{\prime}) \right]
\delta\left( \hat{A} + \frac{A^{\prime}}{1 + A^{\prime}} \right)
\delta\left( \hat{H} - \frac{H^{\prime}}{1 + A^{\prime}} \right), \label{cavity2-Regular}
\end{align}
and
\begin{align}
q_{r}(A, H)
=  \int \prod^{c-1}_{g=1}\left( d\hat{A}_{g} d\hat{H}_{g} \hat{q}_{r}(\hat{A}_{g}, \hat{H}_{g}) \right)
\delta\left( H + \psi/2 - \sum^{c-1}_{g=1}\hat{H}_{g} \right)
\delta\left( A - \phi + \sum^{c-1}_{g=1}\hat{A}_{g} \right). \label{cavity3-Regular}
\end{align}
Note that we set $c > 2$ here. To obtain nontrivial random regular graphs, the degree $c$ of each vertex needs to be greater than $2$.
Furthermore, the saddle-point conditions with respect to the auxiliary parameters $\psi$ and $\phi$ give
\begin{align}
&\sum_{r} p_{r} \int dA dH \, Q_{r}(A,H) \frac{H}{A} = 0, \label{Qr-orthogonality-RegularRandom} \\
&\sum_{r} p_{r} \int dA dH \, Q_{r}(A,H) \left(\frac{H}{A}\right)^{2} = 1, \label{Qr-normalization-RegularRandom}
\end{align}
respectively, where we have defined
\begin{align}
Q_{r}(A,H) = \int \prod^{c}_{g=1}\left( d\hat{A}_{g} d\hat{H}_{g} \hat{q}_{r}(\hat{A}_{g}, \hat{H}_{g}) \right)
\delta\left( H + \psi/2 - \sum^{c}_{g=1}\hat{H}_{g} \right)
\delta\left( A - \phi + \sum^{c}_{g=1}\hat{A}_{g} \right), \label{MarginalDistribution}
\end{align}
which corresponds to the complete marginal of Eq.~(\ref{cavity3-Regular}).
The distribution of $A$ and $H$ in module $r$, $Q_{r}(A, H)$, can be obtained by iteratively updating the saddle-point equations (\ref{cavity1-Regular})--(\ref{cavity3-Regular}), while keeping the constraints (\ref{Qr-orthogonality-RegularRandom}) and (\ref{Qr-normalization-RegularRandom}).

\begin{figure}[t]
\centering
\includegraphics[width=0.8\columnwidth]{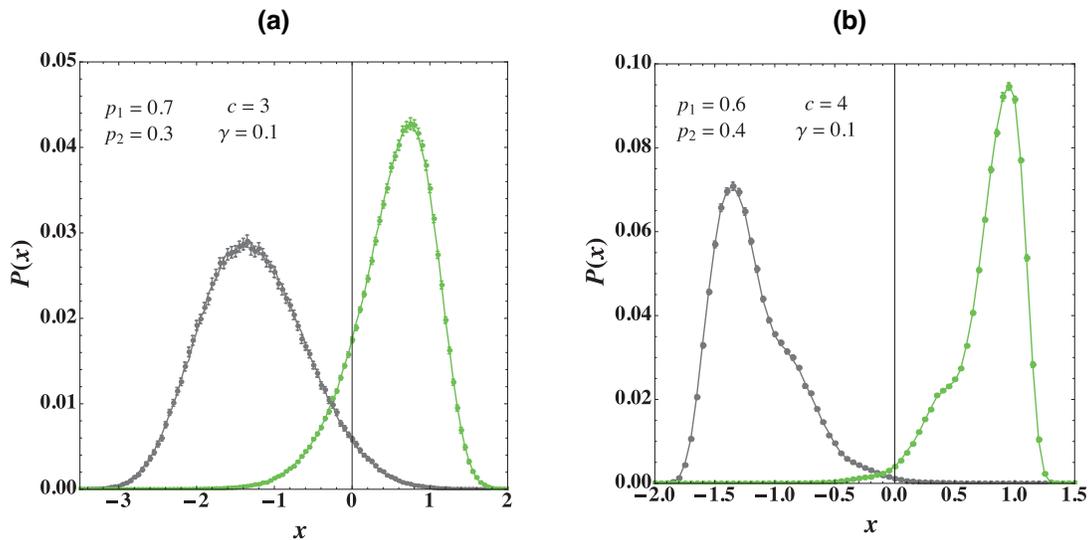}
\caption{
(Color online) Distributions of the elements of the second-smallest eigenvector of the two-block random regular graphs.
The dots in each plot represent the numerical results with $N=10^{4}$ vertices, in which the average is over $100$ samples.
The solid lines in each plot represent the results from the saddle-point equations (\ref{cavity1-Regular})--(\ref{cavity3-Regular}) with parameters (a) $c=3$, $p_{1} = 0.7$, $p_{2} = 0.3$, $\gamma = 0.1$, and (b) $c=4$, $p_{1} = 0.6$, $p_{2} = 0.4$, $\gamma = 0.1$.
}
\label{RegularRandomSecondEigenVector}
\end{figure}

Recall that $\phi$ leads to the normalization condition $\sum_{i} x^{2}_{i}/N = 1$ (see Eq.~(\ref{NormalizationConstraint})). Then, Eq.~(\ref{Qr-normalization-RegularRandom}) should indicate the same restriction, as it incorporates the average over the random graphs at the same time.
Hence, Eq.~(\ref{MarginalDistribution}) indicates that the distribution of $H/A$ gives the distribution $P_{r}(x) = \sum_{i \in V_{r}} \delta(x - x_{i})/N_{r}$ of the elements of the second-smallest eigenvector of each module.
As shown in Fig.~\ref{RegularRandomSecondEigenVector}, the distributions obtained by iterating the saddle-point equations agree very well with the numerical experiments.

\begin{figure}[t]
\centering
\includegraphics[width=0.8\columnwidth]{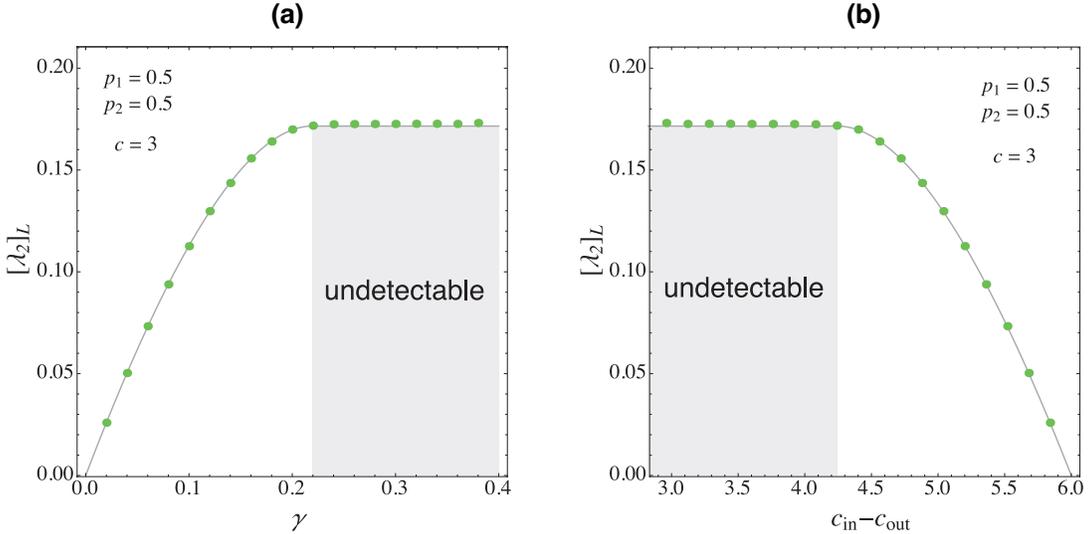}
\caption{ (Color online) Average second-smallest eigenvalues of the unnormalized Laplacian $L$ of the random regular graphs as a function of (a) $\gamma$ and (b) $c_{\mathrm{in}} - c_{\mathrm{out}}$. The solid lines represent the analytical solutions [(\ref{hatm1k-nonzero-solution}) and (\ref{hatm1k-zero-solution})] and the dots represent the numerical results. The numerical experiments used $N=10^{4}$ vertices, and each dot represents the average over $100$ samples.
}
\label{RegularRandomSecondEigenvalues}
\end{figure}

To obtain an analytical expression, we further restrict the form of the solution. We assume that the variances in Eqs.~(\ref{assumedQ}) and (\ref{assumedhatQ}) have the same values $a$ and $\hat{a}$, i.e., $q(A) = \int dH q(A,H) = \delta(a - A)$ and $\hat{q}(\hat{A}) = \int d\hat{H} \hat{q}(\hat{A},\hat{H}) = \delta(\hat{a} - \hat{A})$.
Then, Eq.~(\ref{2ndEigenValue}) becomes
\begin{align}
\left[\lambda_{2}\right]_{L}
&= - \mathrm{extr} \Biggl\{
\frac{1}{a(a+2)} \biggl( c \, p_{1} \left[ (1+a)m_{21} + m_{11}^{2} \right] + c \, p_{2} \left[ (1+a)m_{22} + m_{12}^{2} \right] - \gamma \left( m_{11} - m_{12} \right)^{2} \biggr) \nonumber\\
&\hspace{50pt} -\frac{1}{a-\hat{a}} \sum_{r=1,2} c \, p_{r} \left( m_{2r} + 2m_{1r}\hat{m}_{1r} + \hat{m}_{2r} \right) \nonumber\\
&\hspace{50pt} +\frac{1}{\phi - c \, \hat{a}} \sum_{r=1,2} p_{r} \left( \frac{\psi^{2}}{4} - c \, \psi \hat{m}_{1r} + c \, \hat{m}_{2r} + c (c-1) \hat{m}_{1r}^{2} \right)
+ \phi \Biggr\}, \label{2ndEigenValueFixedA}
\end{align}
where we have denoted the moments of $H$ and $\hat{H}$ as $m_{nr} = \int dH H^{n} q_{r}(H)$ and $\hat{m}_{nr} = \int d\hat{H} \hat{H}^{n} \hat{q}_{r}(\hat{H})$.
Equation (\ref{2ndEigenValueFixedA}) has solutions with $m_{1r} \ne 0$ and $m_{1r} = 0$.
For the solution with $m_{1r} \ne 0$, after taking the saddle point, we have
\begin{align}
&1 + \hat{a} = \frac{1}{1+a}, \label{saddlepoint-hata-RegularRandom}\\
&a = \phi - (c - 1)\hat{a}, \label{saddlepoint-a-RegularRandom}\\
&\hat{m}_{11}^{2}
= \frac{p_{2}}{c p_{1}} \left( 1 - \frac{1}{(c-1)^{2}\Gamma^{2}} \right) \left( (c-1)\Gamma^{2} - 1 \right), \\
&\left[\lambda_{2}\right]_{L}
= (1-\Gamma) \left(c-1-\frac{1}{\Gamma}\right). \label{hatm1k-nonzero-solution}
\end{align}
The requirement that $\hat{m}_{11}^{2} \ge 0$ implies that the above solution is valid for
\begin{align}
\frac{1}{\sqrt{c-1}} \le \Gamma, \label{Detectability-0}
\end{align}
or, in terms of $\gamma$,
\begin{align}
&\gamma \le c f(c) p_{1} p_{2}, \hspace{20pt} \left( f(c) = 1- \frac{1}{\sqrt{c-1}} \right). \label{Detectability-1}
\end{align}
The point at which the equality holds in Eq.~(\ref{Detectability-1}) is the detectability threshold of the random regular graphs. Above this point, we have the solution with $\hat{m}_{1k} = 0$.
When $\hat{m}_{1k} = 0$, we have
\begin{align}
\left[\lambda_{2}\right]_{L}
&= - \phi = c - 2 \sqrt{c - 1}, \label{hatm1k-zero-solution}
\end{align}
which matches Eq.~(\ref{hatm1k-nonzero-solution}) at the boundary of (\ref{Detectability-1}).
In both cases, we have $\psi = 0$, which comes from the symmetry property in which the problem is invariant under conversion from $\bm{x}$ to $-\bm{x}$.
Equations (\ref{hatm1k-nonzero-solution}) and (\ref{hatm1k-zero-solution}) are plotted in Fig.~\ref{RegularRandomSecondEigenvalues}(a), together with the results of the numerical experiments.
Again, the agreement is quite good.

To compare our results with those reported in the literature, we recast Eq.~(\ref{Detectability-0}) in terms of $c_{\mathrm{in}} = p_{\mathrm{in}} N$ and $c_{\mathrm{out}} = p_{\mathrm{out}}N$, and set the module sizes to $p_{1} = p_{2} = 0.5$.
First, we recast Eq.~(\ref{Detectability-1}) in terms of the total degree within a module $K_{\mathrm{in}}$ and the total degree from one module to the other $K_{\mathrm{out}}$.
With these quantities, we have $K_{\mathrm{out}} = \gamma N$ and $K = c N = 2 (K_{\mathrm{in}} + K_{\mathrm{out}})$ for the total degree of the whole graph $K$.
Then, Eq.~(\ref{Detectability-1}) reads
\begin{align}
K_{\mathrm{in}} - K_{\mathrm{out}} &\ge \frac{N}{2} \frac{c}{\sqrt{c-1}}. \label{Detectability-2}
\end{align}
In the limit $N\rightarrow\infty$, $K_{\mathrm{in}} (= cN/2 - \gamma N)$ and $K_{\mathrm{out}} (= \gamma N)$ are related to $c_{\mathrm{in}}$ and $c_{\mathrm{out}}$ as Eqs.~(\ref{Kout}) and (\ref{Kin}), respectively. 
Therefore, the detectable region (\ref{Detectability-2}) is
\begin{align}
c_{\mathrm{in}} - c_{\mathrm{out}} &\ge 2 \frac{c}{\sqrt{c-1}}. \label{Detectability-3}
\end{align}
In the dense limit $c \rightarrow\infty$, Eq.~(\ref{Detectability-3}) converges to the result in \cite{Nadakuditi2012}.
The implication of Eq.~(\ref{Detectability-3}) is that, even if the effect of localization is absent, the spectral method for sparse graphs never reaches the ultimate limit $2 \sqrt{c}$ \cite{Nadakuditi2012,Decelle2011,Mossel2014}.

\begin{figure}[t]
\centering
\includegraphics[width=0.8\columnwidth]{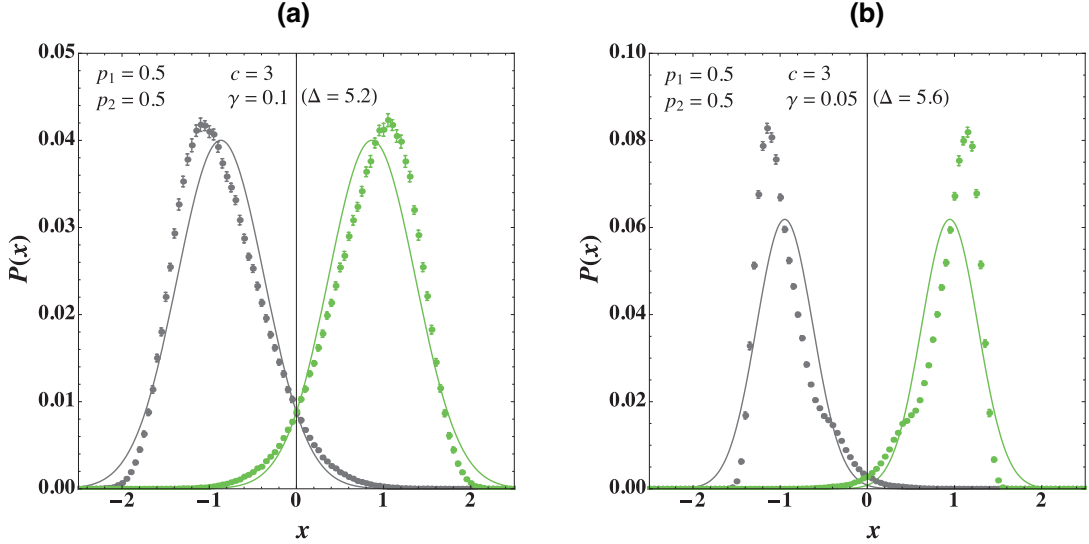}
\caption{
(Color online) Gaussian fitting of the distributions of elements for the second-smallest eigenvector of the two-block $3$-random regular graphs.
The values of $\gamma$ are (a) $0.1$ and (b) $0.05$, and the equal module sizes are considered in both plots.
The dots in each plot represent the results of the numerical experiments with $N=10^{4}$ vertices, in which the average is over $100$ samples.
The solid lines in each plot represent the Gaussian distributions with the same mean and variance as the values in the numerical experiments.
}
\label{RegularRandomGaussianFitting}
\end{figure}

\begin{figure}[t]
\centering
\includegraphics[width=0.5\columnwidth]{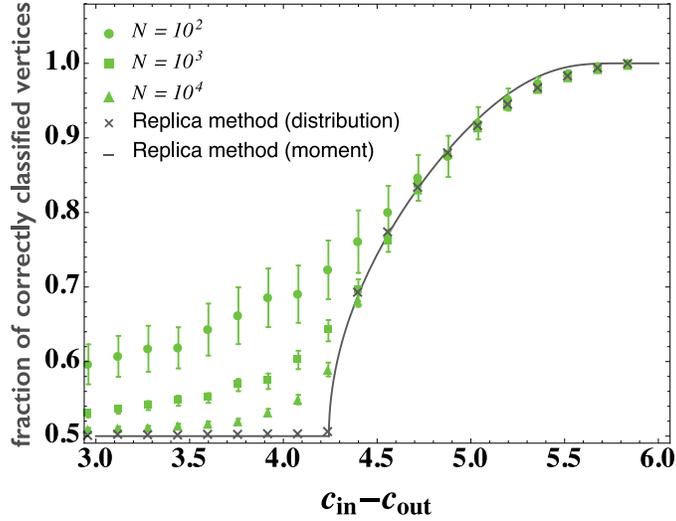}
\caption{(Color online)
Estimate of the fraction of correctly classified vertices in the two-block $3$-random regular graphs with equal module sizes and the numerical results.
The crosses represent the estimates using the replica method, and the solid line represents their Gaussian approximation.
The numerical experiments were conducted for various graph sizes. 
}
\label{CorrectFractionRegularRandom}
\end{figure}

Although the distribution $P_{r}(x)$ is not of Gaussian form, even for the case of equal size modules [see Figs.~\ref{RegularRandomGaussianFitting}(a) and \ref{RegularRandomGaussianFitting}(b) for the apparent cases], it is expected to be somewhat close to the Gaussian distribution, especially when the peaks are not well-separated.
Estimating the mean $m(x)$ and  variance $s^{2}(x)$ from the replica analysis above, we can calculate the mean value of the fraction of correctly classified vertices as
$[1 + \mathrm{erf}(|m(x)| / \sqrt{2 s^{2}(x)}) ]/2$, under the Gaussian approximation (see Appendix~\ref{Appendix-CorrectFraction} for details).
Figure \ref{CorrectFractionRegularRandom} illustrates this estimate, together with the results of numerical experiments and the populations obtained by iterating the saddle-point equations.
The Gaussian fitting appears to give a fairly good approximation, especially around the detectability threshold.
We can also observe some convergence of the numerical results to our estimate, which is for the infinite-size limit.

Note that there exists a finite fraction of misclassified vertices, even in the limit $N \rightarrow\infty$.
Although the poor performance of optimization algorithms may have crucial effects in practice, misclassification occurs in principle  because the planted partition is not necessarily the partition that optimizes the objective function.
Unless the block structure of the planted partition is sufficiently strong, the random graph is likely to have a partition that is better, in the sense of the objective function, than the planted partition.

\begin{figure}[t]
\centering
\includegraphics[width=0.6\columnwidth]{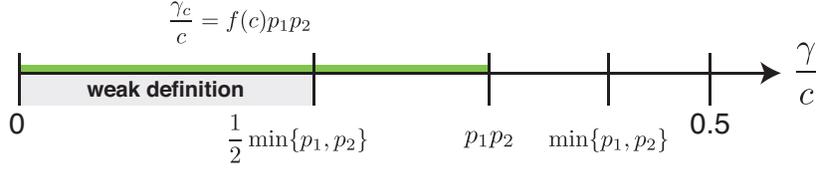}
\caption{(Color online) Parameter region of the detectability threshold for random regular graphs, $f(c)p_{1}p_{2}$, in the space of $\gamma/c$. This is in the range $(0,p_{1}p_{2})$, whereas the region in which the weak definition of a community is satisfied is below $\min\{p_{1}, p_{2}\}/2$.
}
\label{PhysicalInterpretation}
\end{figure}

We now discuss the behavior of the detectability threshold (\ref{Detectability-1}).
We consider $\gamma/c$, the fraction of connections between modules normalized by the degree.
The upper bound of this quantity is $\min\{ p_{1}, p_{2}\}$, which is achieved when all edges in a smaller module are connected to the other module, i.e., the case where the small module has a bipartite structure.
For a uniform random graph, i.e., a graph with no block structures, the expected value of $\gamma/c$ is $p_{1}p_{2}$.
This is because, for each stub or half-edge in the first module in $c N_{1}$, the probability of being connected to the second module is $N_{2}/N$.
Therefore, we have $\gamma N = c N_{1} \times N_{2}/N$.
This can also be obtained as the value of $\gamma$ with $\eta = 0$ in (\ref{etagraph}).
Note that the parameter $1-\Gamma$ is the ratio of $\gamma$ to $cp_{1}p_{2}$.
Because $0<f(c)<1$, the detectability threshold $\gamma_{c}/c$ falls somewhere in the range $(0,p_{1}p_{2})$ (see Fig.~\ref{PhysicalInterpretation}).

Let us consider the value of $\gamma/c$ such that the total degree within a module is greater than the number of edges between modules.
As the number of edges within a smaller module is $cN\min\{p_{1}, p_{2}\} - \gamma N$, we have
\begin{align}
\frac{\gamma_{\mathrm{wd}}}{c} < \frac{1}{2} \min\{p_{1}, p_{2}\}. \label{WeakDefinition}
\end{align}
This corresponds to the region where the weak definition of a community \cite{Radicchi2004} is satisfied with respect to the smaller module.
While the detectability threshold (\ref{Detectability-1}) is always in the region stated in (\ref{WeakDefinition}) for equal-size modules, this may not be the case for unequal-size modules, because the larger module may possess a strong block structure even if the smaller one does not.
Such a situation is achieved when $2 f(c) \max\{p_{1}, p_{2}\}>1$.
In other words, because $f(5) = 1/2$, the spectral method always loses all information about the planted solution when the weak definition is not satisfied in any module size for $c \le 5$.
Note that this is for bisection using the sign of the eigenvector with the second-smallest eigenvalue. Although this is the standard approach, as we mentioned at the end of Sec.~\ref{SpectralMethods}, its performance is not reliable in practice when the module sizes are very different.


\section{Detectability threshold in random graphs with degree fluctuation: the unnormalized Laplacian} \label{DetectabilityRatioCut}
We now analyze the case where the degree fluctuates in the unnormalized Laplacian $L$.
Within $N_{r} = p_{r}N$ vertices,
we consider a graph in which $b_{t} N_{r}$ ($t \in \{1,2, \dots, T \}$, $\sum_{t=1}^{T} b_{t} = 1$) vertices have degree $c_{t}$.
In the limit $N \rightarrow \infty$, the replica-symmetric solution of the second eigenvalue is given by
\begin{align}
\left[\lambda_{2}\right]_{L}
&= - \mathop{\mathrm{extr}}_{q_{r}, \hat{q}_{r}, \phi, \psi} \Biggl\{ \int dA dH \int dA^{\prime} dH^{\prime} \, \Xi(A, H, A^{\prime}, H^{\prime}) \nonumber\\
&\hspace{10pt} \times \frac{1}{2} \biggl(
(\overline{c} \, p_{1} - \gamma) q_{1}(A, H) q_{1}(A^{\prime}, H^{\prime}) +
(\overline{c} \, p_{2} - \gamma) q_{2}(A, H) q_{2}(A^{\prime}, H^{\prime}) +
2 \gamma q_{1}(A, H) q_{2}(A^{\prime}, H^{\prime})
\biggr)
\nonumber\\
&\hspace{10pt} +\phi - \overline{c} \sum_{r = 1,2} p_{r} \int dA dH \int d\hat{A} d\hat{H} \,
q_{r}(A, H) \hat{q}_{r}(\hat{A}, \hat{H})
\left(
\frac{(H + \hat{H})^{2}}{A - \hat{A}} - \frac{H^{2}}{A}
\right) \nonumber\\
&\hspace{20pt} + \sum_{r=1,2} p_{r} \sum_{t} b_{t} \int \prod_{g=1}^{ c_{t} } \left( d\hat{A}_{g}d\hat{H}_{g} \hat{q}_{r}(\hat{A}_{g}, \hat{H}_{g}) \right)
\frac{\left(\psi/2 - \sum_{g}\hat{H}_{g}\right)^{2}}{\phi - \sum_{g}\hat{A}_{g}}
\Biggr\}. \label{distributed-2ndEigenValue}
\end{align}
The saddle-point equations derived from (\ref{distributed-2ndEigenValue}) are
\begin{align}
\hat{q}_{1}(\hat{A}, \hat{H})
= \int dA^{\prime} dH^{\prime} \, \left[ \left( 1 - \frac{\gamma}{\overline{c} \, p_{1}}\right) q_{1}(A^{\prime}, H^{\prime}) + \frac{\gamma}{\overline{c} \, p_{1}} q_{2}(A^{\prime}, H^{\prime}) \right]
\delta\left( \hat{A} + \frac{A^{\prime}}{1 + A^{\prime}} \right)
\delta\left( \hat{H} - \frac{H^{\prime}}{1 + A^{\prime}} \right), \label{cavity1-RatioCut}\\
\hat{q}_{2}(\hat{A}, \hat{H})
= \int dA^{\prime} dH^{\prime} \, \left[ \left( 1 - \frac{\gamma}{\overline{c} \, p_{2}}\right) q_{2}(A^{\prime}, H^{\prime}) + \frac{\gamma}{\overline{c} \, p_{2}} q_{1}(A^{\prime}, H^{\prime}) \right]
\delta\left( \hat{A} + \frac{A^{\prime}}{1 + A^{\prime}} \right)
\delta\left( \hat{H} - \frac{H^{\prime}}{1 + A^{\prime}} \right), \label{cavity2-RatioCut}
\end{align}
and
\begin{align}
q_{r}(A, H)
= \sum_{t} \frac{b_{t}c_{t}}{\overline{c}} \int \prod^{c_{t}-1}_{g=1}\left( d\hat{A}_{g} d\hat{H}_{g} \hat{q}_{r}(\hat{A}_{g}, \hat{H}_{g}) \right)
\delta\left( H + \frac{\psi}{2} - \sum^{c_{t}-1}_{g=1}\hat{H}_{g} \right)
\delta\left( A - \phi + \sum^{c_{t}-1}_{g=1}\hat{A}_{g} \right). \label{cavity3-RatioCut}
\end{align}
Compared with the saddle-point equations for the random regular graphs, we now have a step in (\ref{cavity3-RatioCut}) to draw a degree $c_{t}$ from the excess degree distribution $b_{t}c_{t}/\overline{c}$ at every iteration.

Note that, in the case of the random regular graphs, we had a set of solutions $q_{r}(A,H)$ and $\hat{q}_{r}(A,H)$ of the saddle-point equations (\ref{cavity1-Regular})--(\ref{cavity3-Regular}), with $A$ and $\hat{A}$ fixed at certain values $a$ and $\hat{a}$.
Although these give exact solutions with random regular graphs, functions of this form cannot solve Eqs.~(\ref{cavity1-RatioCut})--(\ref{cavity3-RatioCut}) because of the degree fluctuation.
Although the saddle-point equations (\ref{cavity1-RatioCut})--(\ref{cavity3-RatioCut}) no longer have solutions of simple form, it is still important to obtain an analytical expression.
For this purpose, we again employ $q_{r}(A)$ and $\hat{q}_{r}(\hat{A})$ of the form $q_{r}(A)=\delta(a-A)$ and $\hat{q}_{r}(\hat{a} - \hat{A})$, respectively. This is called the \textit{effective medium approximation} (EMA) \cite{Biroli1999,Kabashima2012}.
We can then recast Eq.~(\ref{distributed-2ndEigenValue}) as
\begin{align}
\left[\lambda_{2}\right]_{L}
= - \mathrm{extr} \Biggl\{
&\phi + \frac{1}{a (a+2)} \left( \overline{c} \sum_{r} p_{r} \left[ (1+a) m_{2r} + m_{1r}^{2} \right] -\gamma (m_{11} - m_{12})^{2} \right) \nonumber\\
&-\overline{c} \, \sum_{k} \frac{p_{r}}{a-\hat{a}} \left( m_{2r} + 2m_{1r}\hat{m}_{1r} + \hat{m}_{2r} \right) \nonumber\\
&+ \sum_{r=1,2} p_{r} \sum_{t} \frac{b_{t}}{\phi - c_{t} \hat{a}} \left( \frac{\psi^{2}}{4} + c_{t}(\hat{m}_{2r} - \psi \hat{m}_{1r}) + c_{t}(c_{t}-1)\hat{m}_{1r}^{2} \right)
\Biggr\}. \label{distributed-2ndEigenValue-EMA}
\end{align}
Introducing
\begin{align}
& R_{n} = \sum_{t} \frac{b_{t} c_{t}^{n}}{\phi - c_{t} \hat{a}}, \\
& S_{n} = \sum_{t} \frac{b_{t} c_{t}^{n}}{(\phi - c_{t} \hat{a})^{2}},
\end{align}
and provided that $a$ and $\phi$ are obtained at the saddle point, we have
\begin{align}
m_{11}^{2} &= \frac{X_{1} - S_{2}}{(S_{2} - S_{1})X_{1}X_{2} + (S_{1}S_{3} - S_{2}^{2})X_{2} - S_{1}X_{3}}, \label{m11-EMA-formal}\\
\left[\lambda_{2}\right]_{L} &= \frac{p_{1}}{p_{2}} m_{11}^{2} \left[ \frac{\overline{c}}{a(a+2)} \overline{\Gamma}
- \frac{R_{2} - R_{1}}{(1+a)^{2}}\overline{\Gamma}^{2} \right] - \phi, \label{2ndEigenValue-RatioCutEMA-formal}
\end{align}
where we set
\begin{align}
& \overline{\Gamma} = 1-\frac{\gamma}{\overline{c} p_{1}p_{2}}, \\
& X_{1} = \frac{R_{1}^{2}}{\overline{c}} (a^{2} + 2a + 2), \\
& X_{2} 
= \frac{p_{1}}{p_{2}} \left(\frac{\overline{\Gamma}}{1+a}\right)^{2}, \\
& X_{3} 
= 2 \frac{R_{1}^{2}}{\overline{c}} \frac{p_{1}}{p_{2}} \frac{\overline{\Gamma}}{1+a}.
\end{align}
We have $m_{11} = 0$ when $S_{2} \ge X_{1}$, i.e.,
\begin{align}
(1+a)^{2} \le \frac{\overline{c}S_{2}}{R_{1}^{2}} - 1. \label{transitionpoint-RatioCutEMA-formal}
\end{align}
Equation (\ref{2ndEigenValue-RatioCutEMA-formal}) gives the formal solution for the second-smallest eigenvalue, and the equality condition in (\ref{transitionpoint-RatioCutEMA-formal}) is our estimate of the detectability threshold with the EMA.

When the graph has a bimodal distribution, i.e., $\{ b_{1}, b_{2}\}$ for $c_{1}$ and $c_{2}$, we can solve for $a$ and $\phi$ at the saddle point analytically.
In this case, the saddle-point conditions give
\begin{align}
\phi = \frac{c_{1}c_{2}a}{1+a} \left( \frac{1+a}{\overline{\Gamma}} + 1 - \overline{c} \right) \left[ \overline{c^{2}} - \overline{c}\left(1 + \frac{1+a}{\overline{\Gamma}} \right) \right]^{-1}, \label{phiop}
\end{align}
where we have defined $\overline{c^{2}} = b_{1} c_{1}^{2} + b_{2} c_{2}^{2}$, and
$a$ is the solution of the following quadratic equation:
\begin{align}
\overline{c} \left( \frac{1+a}{\overline{\Gamma}} \right)^{2}
+ \left( \overline{c} - \overline{c^{2}} + \frac{c_{1}c_{2}}{\overline{\Gamma}-1} \right) \frac{1+a}{\overline{\Gamma}}
+ \frac{c_{1}c_{2}(1-\overline{c})}{\overline{\Gamma}-1} = 0. \label{aop}
\end{align}
We take the smaller value for the solution of Eq.~(\ref{aop}), which gives a nonnegative value for $m_{11}^{2}$.

\begin{figure}[!t]
\centering
\includegraphics[width=0.8\columnwidth]{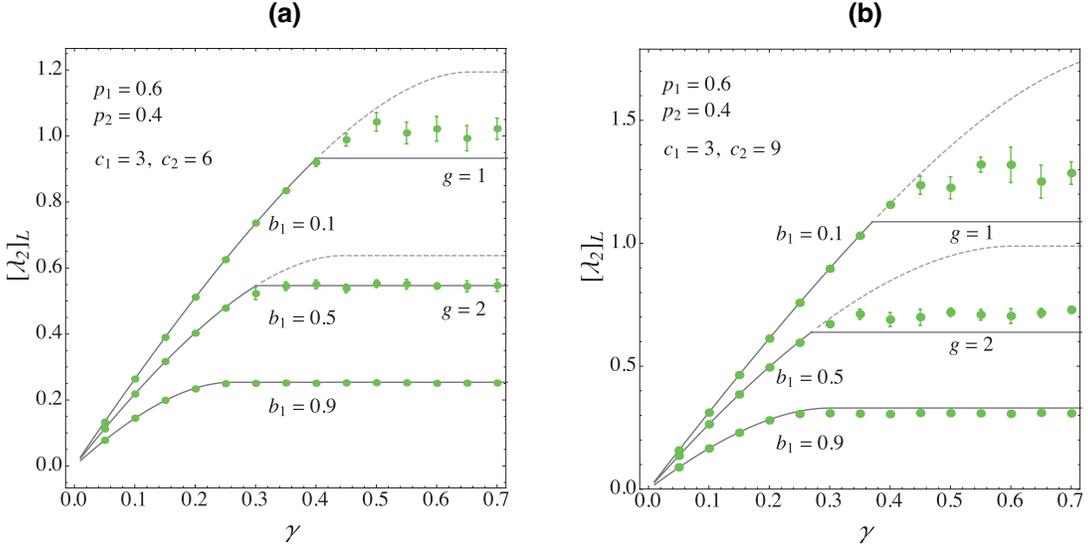}
%
%
%
    \caption{(Color online)
    Average second-smallest eigenvalues of the unnormalized Laplacian $L$ of the two-block random graphs with bimodal degree distributions, as a function of $\gamma$.
    The degree sets are (a) $\{c_{1}, c_{2}\} = \{3, 6\}$ and (b) $\{c_{1}, c_{2}\} = \{3, 9\}$.
    The ratio of module sizes is set to $p_{1} = 0.6$ ($p_{2} = 0.4$) in both cases.
    The estimates given by the replica method with the EMA and the estimated eigenvalues for the localized eigenvectors are represented by the solid lines (see Sec.~\ref{LocalizationRatioCut} for details of the localized eigenvectors).
    When the latter gives a lower eigenvalue, the former is indicated by a dashed line.
    The dots represent the  numerical results with $N = 10^{4}$. The average is over ten samples. In each plot, results are plotted, from top to bottom, for $b_{1} = 0.1, 0.5, 0.9$ ($b_{2} = 0.9, 0.5, 0.1$).}
    \label{RatioCutEMAEigenvalues}
\end{figure}

Figures~\ref{RatioCutEMAEigenvalues}(a) and \ref{RatioCutEMAEigenvalues}(b) show the eigenvalues of the unnormalized Laplacian with the EMA and those of the localized eigenvectors, together with the results of the numerical experiments.
The estimate of the localized eigenvectors and their eigenvalues is discussed in Sec.~\ref{LocalizationRatioCut}, and we describe the resulting behavior here.
As $\gamma$ increases, the eigenvector possessing information about the modules will eventually have a higher eigenvalue than that of the localized eigenvector; the standard spectral method fails in such a region.
Our estimate agrees very well with the numerical result, as long as localization does not occur.
Although we do not  know which $g$ should be chosen \textit{a priori} (see Sec.~\ref{LocalizationRatioCut} for the meaning of $g$), if we choose a value that is consistent with the result of the numerical experiment, our estimate of the localization transition is close to the point where the result given by the replica method with the EMA starts to deviate from numerical result.

Eigenvector localization is expected to result from the existence of a few vertices with irregular degrees, which we call the \textit{defects}.
In Figs.~\ref{RatioCutEMAEigenvalues}(a) and \ref{RatioCutEMAEigenvalues}(b), we regard the vertices with the lower population as the defects.
As we show in Secs.~\ref{LocalizationRatioCut} and \ref{LocalizationNcut}, localization tends to occur in both unnormalized and  normalized Laplacians when the defects have a lower degree.
Therefore, when we have equal populations, i.e., $b_{1} = b_{2} = 0.5$, we regard vertices with lower degree as defects.
Indeed, when vertices with lower degree are dominant, localization does not seem to occur, or produces only a negligible effect.

\section{Localized eigenvector of the unnormalized Laplacian with bimodal degree distributions} \label{LocalizationRatioCut}
The behavior of the eigenvalues of  localized eigenvectors for the unnormalized Laplacian $L$ was described in Sec.~\ref{DetectabilityRatioCut}. This section is devoted to their derivation and an analysis of some specific examples.
As mentioned in the previous section, localized eigenvectors emerge because of degree fluctuations.
They are the vectors in which the weight of their elements is concentrated around a few defects, the vertices with characteristic degrees.
Here, we analyze the localization of an eigenvector for the unnormalized Laplacian $L$ and consider this process for the normalized Laplacian $\mathcal{L}$ in Sec.~\ref{LocalizationNcut}.
We focus on sparse graphs with bimodal degree distributions. The graphs have two types of degree, $c_{D}$ and $c_{B}$, with  populations $b_{D}$ and $b_{B}$ ($b_{D} + b_{B} = 1$), respectively.
We let $b_{D} < b_{B}$ and refer to the vertices with degree $c_{D}$ as the defects.

\begin{figure}[t]
\centering
\includegraphics[width=0.35\columnwidth]{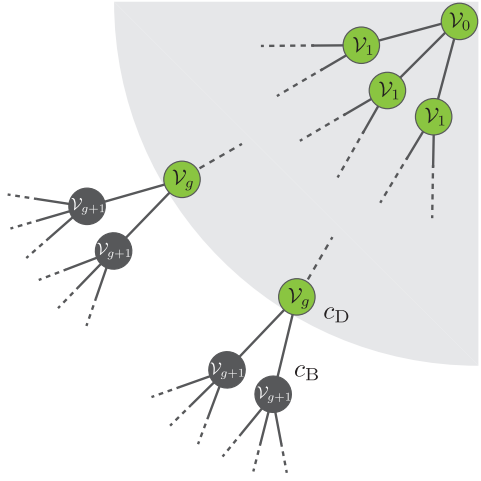}
\caption{(Color online) A tree with the defects aggregated around its root.
}
\label{DefectTree}
\end{figure}

As frequently analyzed for sparse graphs in the literature \cite{Biroli1999,Semerjian2002,Metz2010,Kabashima2012} (see Fig.~\ref{DefectTree}), we consider a tree with defects aggregated around its root, up to a distance $g$ from the root.
Hereafter, we denote a localized eigenvector as $\bm{v}$.
We now simplify the graph  by letting all  vertices at distance $d > g$ have a uniform degree $c_{B}$.
Then, the symmetry of the graph implies that, for all  vertices located at distance $d$ from the root, $v_{i} = \mathcal{V}_{d}$.
From the definition of the eigenvector $L \bm{v} = \lambda \bm{v}$, for $g \ge 1$, we have
\begin{align}
(c_{D} - \lambda) \mathcal{V}_{0} - c_{D} \mathcal{V}_{1} &= 0, \nonumber\\
(c_{D} - 1) \mathcal{V}_{d+1} - (c_{D} - \lambda) \mathcal{V}_{d} + \mathcal{V}_{d-1} &= 0 \hspace{20pt} (0 < d \le g), \nonumber\\
(c_{B} - 1) \mathcal{V}_{d+1} - (c_{B} - \lambda) \mathcal{V}_{d} + \mathcal{V}_{d-1} &= 0 \hspace{20pt} (d \ge g+1). \label{DefectEq-unnormalized}
\end{align}
To be a localized eigenvector, the element $\mathcal{V}_{d}$ needs to vanish at $d \rightarrow \infty$. Thus, we choose a solution of the form $\mathcal{V}_{d+1} = \kappa \mathcal{V}_{d}$ with $|\kappa| < 1$ for $d > g$, where $\kappa$ satisfies
\begin{align}
(c_{B} - 1) \kappa^{2} - (c_{B} - \lambda) \kappa + 1 = 0. \label{CharacteristicEq-unnormalized}
\end{align}
In addition, we have the constraint that the norm of the eigenvector needs to be finite, i.e., $|\bm{v}|^{2} < \infty$.
Let us consider the case $g=0$, i.e., only the vertex at the root is defective.
From Eq.~(\ref{DefectEq-unnormalized}), $\lambda$ must satisfy
\begin{align}
\frac{c_{D} - \lambda}{c_{D}} = \frac{\mathcal{V}_{1}}{\mathcal{V}_{0}} = \frac{\mathcal{V}_{2}}{\mathcal{V}_{1}} = \kappa(\lambda),
\end{align}
where $\kappa(\lambda)$ is a function of $\lambda$ determined by Eq.~(\ref{CharacteristicEq-unnormalized}).
The resulting non-zero eigenvalue is $\lambda = c_{D} (c_{B}-c_{D}-2) / (c_{B}-c_{D}-1)$, and the corresponding damping factor $\kappa$ is $\kappa = 1/(c_{B} - c_{D} - 1)$.
In addition, the constraint finite norm requires $c_{D} < c_{B} -1 -\sqrt{c_{B}-1}$.
Interestingly, the defect must have a lower degree than the other vertices, whereas, in the case of the adjacency matrix, it was hubs that caused  localization.
Note, however, that this is the condition for a tree with uniform degree at $d > 0$ and does not hold in general.

Let us now consider using the results given by the replica method with the EMA from the previous section in the case where the vertex degrees  at $d > g$ are not necessarily uniform.
Recall that $\left[ \lambda_{2} \right]_{L} = - \phi$ when $m_{11} = 0$, and that $\hat{a}$ satisfies Eqs.~(\ref{saddlepoint-hata-RegularRandom}) and (\ref{saddlepoint-a-RegularRandom}) for random regular graphs.
From these relations, we see that the factor $1+\hat{a}$ obeys the same characteristic equation as (\ref{CharacteristicEq-unnormalized}).
This connection can also be seen from the saddle-point equation; 
Eqs. (\ref{cavity1-Regular})--(\ref{cavity3-Regular}) yield $\partial H/\partial H_{g} = (1+a)^{-1} = 1+\hat{a}$. This implies that, in tree-like graphs, the response function $\partial \mathcal{V}_{i} /\partial H_{j}$ generally damps as $O((1+\hat{a})^{D(i,j)})$, where $D(i,j)$ is the distance between sites $i$ and $j$. In particular, taking vertex $j$ to the root ($g=0$) and comparing the relation with the solution form $\mathcal{V}_{d} = const \, \kappa^{d}$ for $d>g$, where $const$ is proportional to $\mathcal{V}_{0}$, we conclude $\kappa=1+\hat{a}$.
Hence, one way to estimate the localized eigenvector and its eigenvalue is to replace the damping factor $\kappa$ with $1 + \hat{a}_{\ast}(\lambda)$, where $\hat{a}_{\ast}(\lambda)$ is the value of $\hat{a}$ obtained via the EMA as a function of $\lambda$.
Then, analogously to the example above, we numerically compute the eigenvalue $\lambda$ so that it is consistent with the equations for $d \le g$ in Eq.~(\ref{DefectEq-unnormalized}), i.e.,
\begin{align}
\frac{\mathcal{V}_{g+1}}{\mathcal{V}_{g}} = \frac{\mathcal{V}_{g+2}}{\mathcal{V}_{g+1}} = 1 + \hat{a}_{\ast}(\lambda). \label{ConjunctionCondition}
\end{align}
If this eigenvalue is smaller than the second-smallest eigenvalue  calculated in the previous section, we can conclude that the localization transition has occurred.

\section{Detectability threshold in random graphs with degree fluctuation: the normalized Laplacian} \label{DetectabilityNcut}
We now analyze the spectral method with the normalized Laplacian $\mathcal{L}$ and consider the two-block random graph with degree fluctuations.
With $\bm{y} := D^{-1/2} \bm{x}$, the Hamiltonian can be written as
\begin{align}
&H(\bm{x} | \mathcal{L}) = \frac{1}{2} \bm{x}^{\mathrm{T}} \mathcal{L} \bm{x}
= \frac{1}{2} \bm{y}^{\mathrm{T}} L \bm{y},
\end{align}
and, noting that the total degree reads $K = \overline{c}N$, we define the partition function in terms of $\bm{y}$ as
\begin{align}
Z(\beta | \mathcal{L})
&= \int d \bm{y} \, \mathrm{e}^{-\beta H(\bm{y} | L)}
\delta(\bm{y}^{\mathrm{T}} D \bm{y} - \overline{c} N)
\delta(\bm{1}^{\mathrm{T}} D \bm{y}), \label{NcutPartitionFunction}
\end{align}
where we have omitted the constant factor obtained by defining the partition function in terms of $\bm{x}$.
The free energy defined by this partition function is related to the second-smallest eigenvalue according to
\begin{align}
2 \lim_{\beta \rightarrow \infty} \left[ f(\beta | \mathcal{L}) \right]_{\mathcal{L}}
&= -2 \lim_{\beta \rightarrow \infty} \frac{1}{N\beta} \left[ \ln Z(\beta | \mathcal{L}) \right]_{\mathcal{L}}
= \overline{c} \left[ \lambda_{2} \right]_{\mathcal{L}}.
\end{align}
Note that the vector $\bm{y}$ is not the eigenvector of $\mathcal{L}$, and therefore the distribution $P(y)$ does not give the distribution of the second-smallest eigenvector.
However, as the sign of each vector element is not changed by the conversion from $\bm{y}$ to $\bm{x}$, the fraction of correctly classified vertices is readily obtained from the distribution $P(y)$.

In the limit $N \rightarrow \infty$ (as in Secs.~\ref{DetectabilityRegularRandom} and \ref{DetectabilityRatioCut}), the replica-symmetric solution can be obtained as
\begin{align}
\overline{c} \left[\lambda_{2}\right]_{\mathcal{L}}
= - \mathop{\mathrm{extr}}_{q_{r}, \hat{q}_{r}, \phi, \psi} \Biggl\{ &\int dA dH \int dA^{\prime} dH^{\prime} \, \Xi(A, H, A^{\prime}, H^{\prime}) \nonumber\\
&\times \frac{1}{2} \biggl(
(\overline{c} \, p_{1} - \gamma) q_{1}(A, H) q_{1}(A^{\prime}, H^{\prime}) +
(\overline{c} \, p_{2} - \gamma) q_{2}(A, H) q_{2}(A^{\prime}, H^{\prime}) +
2 \gamma q_{1}(A, H) q_{2}(A^{\prime}, H^{\prime})
\biggr)
\nonumber\\
&+ \overline{c}\, \phi - \overline{c} \sum_{r = 1,2} p_{r} \int dA dH \int d\hat{A} d\hat{H} \,
q_{r}(A, H) \hat{q}_{r}(\hat{A}, \hat{H})
\left(
\frac{(H + \hat{H})^{2}}{A - \hat{A}} - \frac{H^{2}}{A}
\right) \nonumber\\
&+ \sum_{r=1,2} p_{r} \sum_{t} b_{t} \int \prod_{g=1}^{c_{t}} \left( d\hat{A}_{g}d\hat{H}_{g} \hat{q}_{r}(\hat{A}_{g}, \hat{H}_{g}) \right)
\frac{\left(c_{t} \psi/2 - \sum_{g}\hat{H}_{g}\right)^{2}}{c_{t} \phi - \sum_{g}\hat{A}_{g}}
\Biggr\}. \label{distributed-Ncut-2ndEigenValue}
\end{align}
We have the same saddle-point equations as (\ref{cavity1-RatioCut}) and (\ref{cavity2-RatioCut}) for $\hat{q}_{r}(\hat{A}, \hat{H})$.
For $q_{r}(A, H)$, we have an analogous equation to (\ref{cavity3-RatioCut}), but with $\psi$ and $\phi$ replaced with $c_{t} \psi$ and $c_{t}\phi$, i.e.,
\begin{align}
q_{r}(A, H)
= \sum_{t} \frac{b_{t}c_{t}}{\overline{c}} \int \prod^{c_{t}-1}_{g=1}\left( d\hat{A}_{g} d\hat{H}_{g} \hat{q}_{r}(\hat{A}_{g}, \hat{H}_{g}) \right)
\delta\left( H + \frac{c_{t} \psi}{2} - \sum^{c_{t}-1}_{g=1}\hat{H}_{g} \right)
\delta\left( A - c_{t}\phi + \sum^{c_{t}-1}_{g=1}\hat{A}_{g} \right). \label{cavity3-Ncut}
\end{align}

With the EMA, Eq.~(\ref{distributed-Ncut-2ndEigenValue}) is approximated as
\begin{align}
\left[\lambda_{2}\right]_{\mathcal{L}}
= - \mathrm{extr} \Biggl\{
&\phi + \frac{1}{a (a+2)} \left( \sum_{k} p_{k} \left[ (1+a) m_{2k} + m_{1k}^{2} \right] -\frac{\gamma}{\overline{c}} (m_{11} - m_{12})^{2} \right) \nonumber\\
&- \frac{1}{a-\hat{a}} \sum_{k} p_{k} \left( m_{2k} + 2m_{1k}\hat{m}_{1k} + \hat{m}_{2k} \right) \nonumber\\
&+ \frac{1}{\overline{c}(\phi - \hat{a})} \left[ \frac{\overline{c} \psi^{2}}{4} + \sum_{k=1,2} p_{k}
\left( \hat{m}_{2k} - \overline{c}\psi \hat{m}_{1k} + (\overline{c}-1)\hat{m}_{1k}^{2} \right) \right]
\Biggr\}. \label{Ncut-distributed-2ndEigenValue-EMA}
\end{align}
From the saddle-point conditions, for the solution with $m_{1k} \ne 0$, we have
\begin{align}
& 1 + \hat{a} = \frac{1}{1+a}, \label{saddlepoint-hata-Ncut-EMA}\\
& a = \overline{c} \, \phi - (\overline{c} - 1) \hat{a}, \label{saddlepoint-a-Ncut-EMA}\\
& m_{11}^{2} = (\overline{c}-1)^{2} \frac{p_{2}}{\overline{c}p_{1}} \left( 1 - \frac{1}{(\overline{c}-1)^{2} \overline{\Gamma}^{2}} \right) \left( (\overline{c}-1) \overline{\Gamma}^{2} - 1 \right), \label{Ncut-EMA-m11} \\
& \left[\lambda_{2}\right]_{\mathcal{L}} = - \phi = \frac{1-\overline{\Gamma}}{\overline{c}} \left( \overline{c} - 1 - \frac{1}{\overline{\Gamma}} \right), \label{Ncut-distributed-2ndEigenValue-EMA-2}
\end{align}
and in the region where $m_{1k} = 0$, we have 
\begin{align}
\left[\lambda_{2}\right]_{\mathcal{L}} = - \phi = \frac{ \left( \sqrt{\overline{c}-1} - 1 \right)^{2} }{\overline{c}}. \label{Ncut-undetectable-2ndEigenValue-EMA}
\end{align}
Again,  $\psi = 0$ in both cases.
The condition for the solution with $m_{1k} \ne 0$ to exist is
\begin{align}
\frac{1}{\sqrt{\overline{c}-1}} \le \overline{\Gamma}, \label{NcutDetectability1}
\end{align}
and the detectability threshold is the condition when the equality holds.

\begin{figure}[!t]
\centering
\includegraphics[width=0.8\columnwidth]{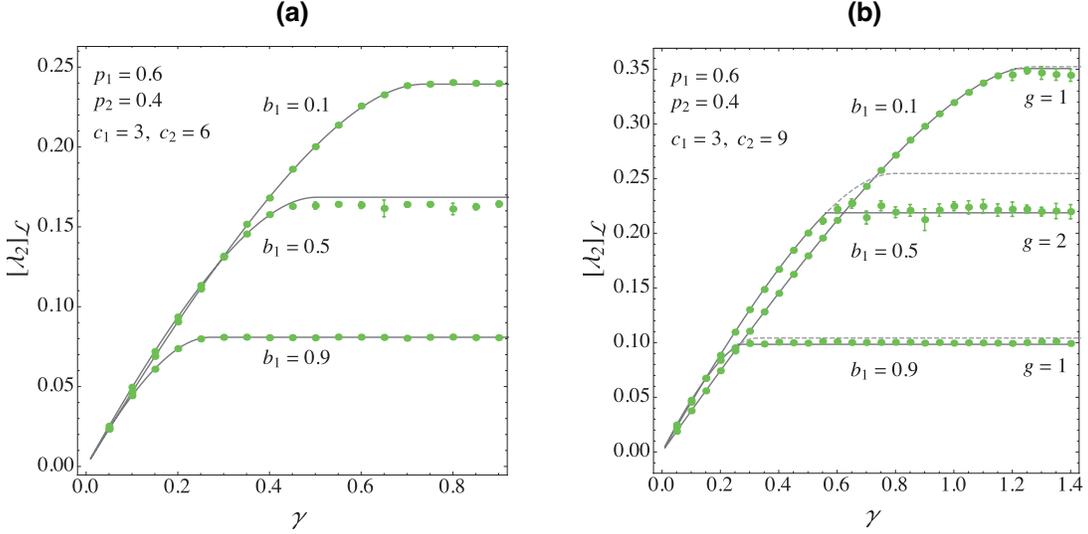}
%
%
%
        \caption{(Color online)
    Average second-smallest eigenvalues of the normalized Laplacian $\mathcal{L}$ of two-block random graphs with bimodal degree distributions as functions of $\gamma$.
    The degree sets are (a) $\{c_{1}, c_{2}\} = \{3, 6\}$ and (b) $\{c_{1}, c_{2}\} = \{3, 9\}$.
    The ratio of module sizes is set to  $p_{1} = 0.6$ ($p_{2} = 0.4$) in both cases.
    The estimates given by the replica method with the EMA and the estimated eigenvalues of the localized eigenvectors are represented by solid lines (see Sec.~\ref{LocalizationNcut} for details of the localized eigenvector).
    When the latter gives a lower eigenvalue, the former is indicated by a dashed line.
    The dots represent the  numerical results with $N = 10^{4}$. The average is over $10$ samples. In each plot, results are plotted, from top to bottom, for $b_{1} = 0.1, 0.5, 0.9$ ($b_{2} = 0.9, 0.5, 0.1$).}
    \label{NcutEMAEigenvalues}
\end{figure}

Figures~\ref{NcutEMAEigenvalues}(a) and \ref{NcutEMAEigenvalues}(b) plot Eqs.~(\ref{Ncut-distributed-2ndEigenValue-EMA-2}) and (\ref{Ncut-undetectable-2ndEigenValue-EMA}), together with the results of the numerical experiments for bimodal distributions.
These figures are plotted in the same way as in Figs.~\ref{RatioCutEMAEigenvalues}(a) and \ref{RatioCutEMAEigenvalues}(b).
Again, our estimate gives a fairly accurate prediction of the  numerical results \cite{CommentNcutBimodal}.
As conjectured in Sec.~\ref{LocalizationNcut}, a comparison of Figs.~\ref{RatioCutEMAEigenvalues}(a) and \ref{RatioCutEMAEigenvalues}(b) and Figs.~\ref{NcutEMAEigenvalues}(a) and \ref{NcutEMAEigenvalues}(b) shows that localization is less likely to occur with the normalized Laplacian $\mathcal{L}$ than with the unnormalized Laplacian $L$.

\begin{figure}[!t]
\centering
\includegraphics[width=0.8\columnwidth]{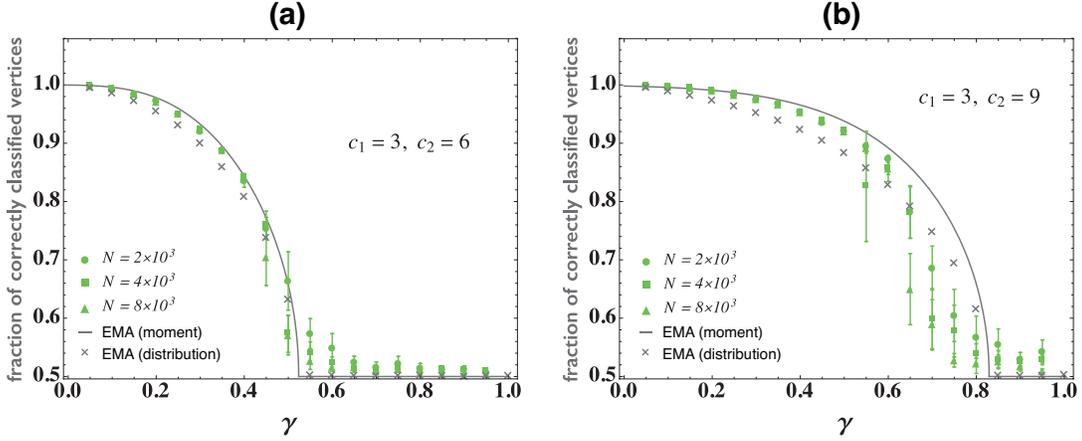}
        \caption{(Color online)
        Fraction of correctly classified vertices with the spectral method for two-block random graphs with bimodal degree distributions as functions of $\gamma$.
        The normalized Laplacian $\mathcal{L}$ is used, and the degree sets are (a) $\{c_{1}, c_{2}\} = \{3, 6\}$ and (b) $\{c_{1}, c_{2}\} = \{3, 9\}$.
    The population of each degree and the ratio of module sizes are $b_{1} = 0.5$ ($b_{2} = 0.5$) and $p_{1} = 0.5$ ($p_{2} = 0.5$) in both cases.
    The dots represent the  numerical results with various graph sizes.
    The crosses represent the results obtained by iterating the saddle-point equations (\ref{cavity1-RatioCut}), (\ref{cavity2-RatioCut}), and (\ref{cavity3-Ncut}).
    The solid lines represent the results obtained by fitting the distribution of eigenvector elements to a Gaussian distribution whose mean and variance are estimated by the saddle point of the free energy with the EMA (see Appendix~\ref{Appendix-CorrectFraction} for details).
    }
    \label{NcutCorrectFractionBimodal}
\end{figure}

Figures \ref{NcutCorrectFractionBimodal}(a) and \ref{NcutCorrectFractionBimodal}(b) show the fraction of correctly classified vertices, ignoring the effect of localization.
Although our estimates are close to the numerical results when the effect of localization is negligible, they differ significantly in the region where  localization is present.
Note that the difference between the crosses and the solid line in each figure is not totally due to the Gaussian approximation of the distribution of the eigenvector elements (see Appendix~\ref{EMAdifference} for details).
In this case, the Gaussian fitting (solid line) whose mean and variance are estimated by the EMA of the free energy seems to be a better approximation.

\section{Localized eigenvector of the normalized Laplacian with bimodal degree distributions} \label{LocalizationNcut}
In this section, we analyze the localized eigenvectors of the normalized Laplacian $\mathcal{L}$.
The behavior of the eigenvalues of these localized eigenvectors for the normalized Laplacian $\mathcal{L}$ were described in Sec.~\ref{DetectabilityNcut}, and this section is devoted to their derivation and an analysis of some specific examples.

We  consider the tree in which the defects are aggregated around its root up to a distance of $g$.
The analysis here is completely analogous to that for the unnormalized Laplacian $L$ in Sec.~\ref{LocalizationRatioCut}.
The equations corresponding to (\ref{DefectEq-unnormalized}) for $g \ge 1$ are
\begin{align}
\mathcal{V}_{1} - (1 - \lambda) \mathcal{V}_{0} &= 0, \nonumber\\
(c_{B}c_{2})^{-1/2}(c_{2} - 1) \mathcal{V}_{g+1} - (1 - \lambda) \mathcal{V}_{g} + c_{2}^{-1} \mathcal{V}_{g-1} &= 0, \nonumber\\
c_{B}^{-1}(c_{B} - 1) \mathcal{V}_{g+2} - (1 - \lambda) \mathcal{V}_{g+1} + (c_{B}c_{2})^{-1/2} \mathcal{V}_{g} &= 0, \nonumber\\
c_{B}^{-1}(c_{B} - 1) \mathcal{V}_{d+1} - (1 - \lambda) \mathcal{V}_{d} + c_{B}^{-1} \mathcal{V}_{d-1} &= 0 \hspace{20pt} (d \ge g + 2). \label{DefectEq-normalized}
\end{align}
As  in Sec.~\ref{LocalizationRatioCut}, we consider the case where the vertices at $d > g$ have a uniform degree $c_{B}$.
Interestingly, when $g=0$, an analogous calculation as for the unnormalized Laplacian $L$ in Sec.~\ref{LocalizationRatioCut} yields that the eigenvector with a non-zero eigenvalue of the form $\mathcal{V}_{d+1} = \kappa \mathcal{V}_{d}$ for $d>1$ has $\kappa = -1$, i.e., we never have a localized eigenvector.
Moreover, when $g=1$, the condition $2 c_{D} < c_{B}$ must be satisfied  to give a localized eigenvector with a finite norm; again, the defects need to have a lower degree.
Note also that the results obtained here are more severe than the conditions in Sec.~\ref{LocalizationRatioCut} for the unnormalized Laplacian $L$ in the sparse case.
This implies that  localization tends to be suppressed in the normalized Laplacian $\mathcal{L}$.
In fact, this tendency is consistent with the analysis in data clustering \cite{Luxburg2007,von2008consistency}.

In the case where the vertex degrees at $d > g$ are not necessarily uniform, a localized eigenvector can be estimated in almost the same way as for the unnormalized Laplacian $L$ in Sec.~\ref{LocalizationRatioCut}, utilizing the results of the replica method with the EMA in Sec.~\ref{DetectabilityNcut}.
Note that, unlike the case of the unnormalized Laplacian $L$, the value of $c_{B}$ is needed to determine the ratio $\mathcal{V}_{g+2}/\mathcal{V}_{g+1}$ in Eq.~(\ref{DefectEq-normalized}).
Thus, when we solve for the consistent eigenvalue by $\mathcal{V}_{g+2}/\mathcal{V}_{g+1} = \mathcal{V}_{g+3}/\mathcal{V}_{g+2} = 1 + \hat{a}_{\ast}(\lambda)$, we replace $c_{B}$ in the ratio $\mathcal{V}_{g+2}/\mathcal{V}_{g+1}$ with the average degree $\overline{c}$.

\section{Detectability threshold of the stochastic block model with the normalized Laplacian} \label{DetectabilitySBM}
\begin{figure}[t]
\centering
\includegraphics[width=0.5\columnwidth]{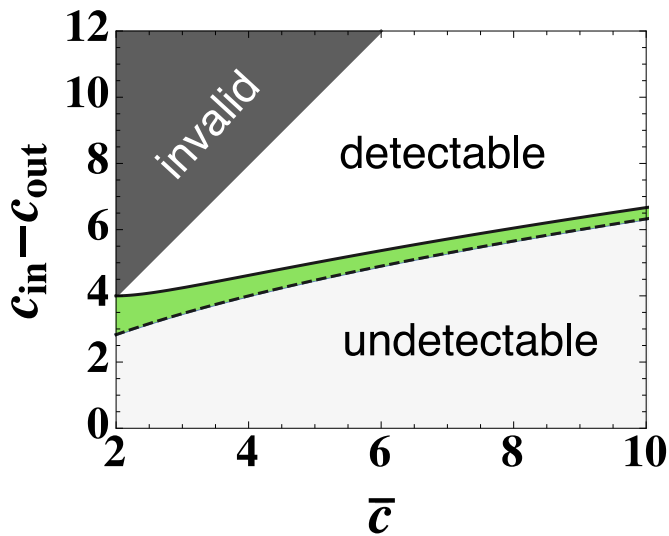}
\caption{
(Color online) Phase diagram of the detectable and undetectable regions given by the replica method with the EMA (solid line) and the ultimate threshold \cite{Nadakuditi2012,Decelle2011,Mossel2014} (dashed line).
The model cannot take parameter values with $c_{\mathrm{in}} - c_{\mathrm{out}} > 2 \overline{c}$ (invalid region) because of the condition that $c_{\mathrm{out}} \ge 0$.
}
\label{ThresholdComparison}
\end{figure}

Finally, we consider the eigenvalues and detectability threshold in the stochastic block model, i.e., the random graph with a Poisson degree distribution.
To compare  with the literature, we recast our result in terms of
$c_{\mathrm{in}}$ and $c_{\mathrm{out}}$ and set the module sizes to $p_{1} = p_{2} = 0.5$.
With these quantities,
the detectability threshold in the normalized Laplacian, Eq.~(\ref{NcutDetectability1}), reads
\begin{align}
&c_{\mathrm{in}} - c_{\mathrm{out}} = \frac{2\overline{c}}{\sqrt{\overline{c}-1}}. \label{NcutDetectability2}
\end{align}
Compared with the threshold obtained in \cite{Nadakuditi2012}, we have a correction factor of $\sqrt{\overline{c}}/\sqrt{\overline{c}-1}$.
The phase diagrams of these two thresholds are shown in Fig.~\ref{ThresholdComparison}.
While the difference between them is negligible when the average degree is sufficiently large, considering the fact that the upper bound of the parameter $c_{\mathrm{in}} - c_{\mathrm{out}}$ is $2 \overline{c}$, this gap is indeed considerable in sparse graphs.

\begin{figure}[!t]
\centering
\includegraphics[width=0.5\columnwidth]{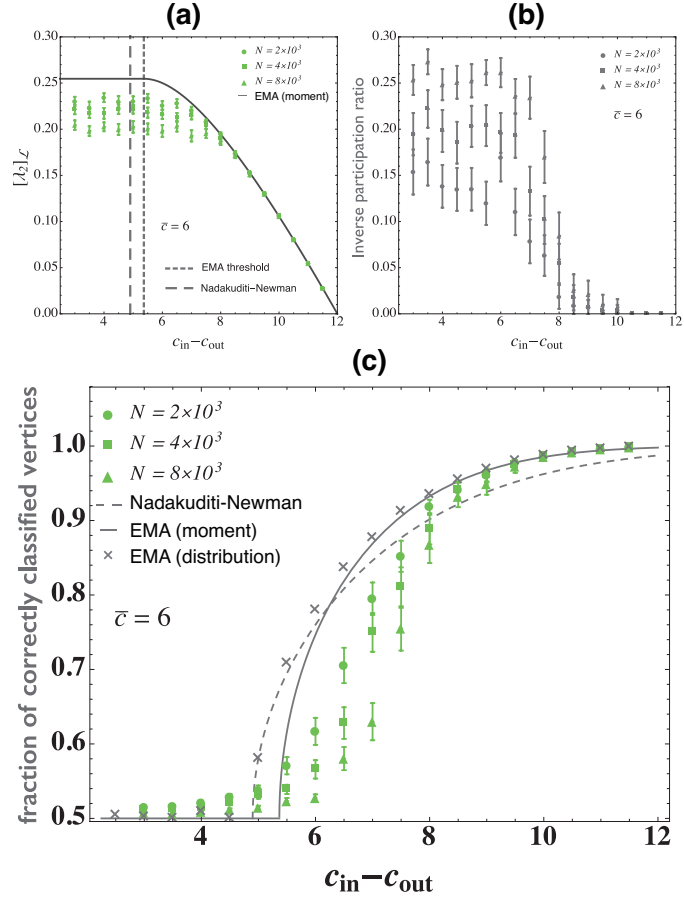}
\caption{
(Color online)
(a) Average second-smallest eigenvalue of the method with the normalized Laplacian $\mathcal{L}$, (b) the IPR of its eigenvector, and (c) the fraction of  correctly classified vertices with the spectral method in the stochastic block model.
We set the average degree $\overline{c} = 6$.
In each plot, the dots represent the numerical results with various graph sizes. 
In (a), the estimate of the eigenvalue with the EMA is represented by a solid line.
The dotted line shows the estimated detectability threshold and the dashed line shows the estimate in \cite{Nadakuditi2012}.
In (c), the crosses represent the fraction of correctly classified vertices with the EMA and the solid line represents their Gaussian approximation (see Appendix~\ref{Appendix-CorrectFraction} for details).
The dashed line is, again, the estimate in \cite{Nadakuditi2012}.
}
\label{SBM-cm6}
\end{figure}

\begin{figure}[!t]
\centering
\includegraphics[width=0.5\columnwidth]{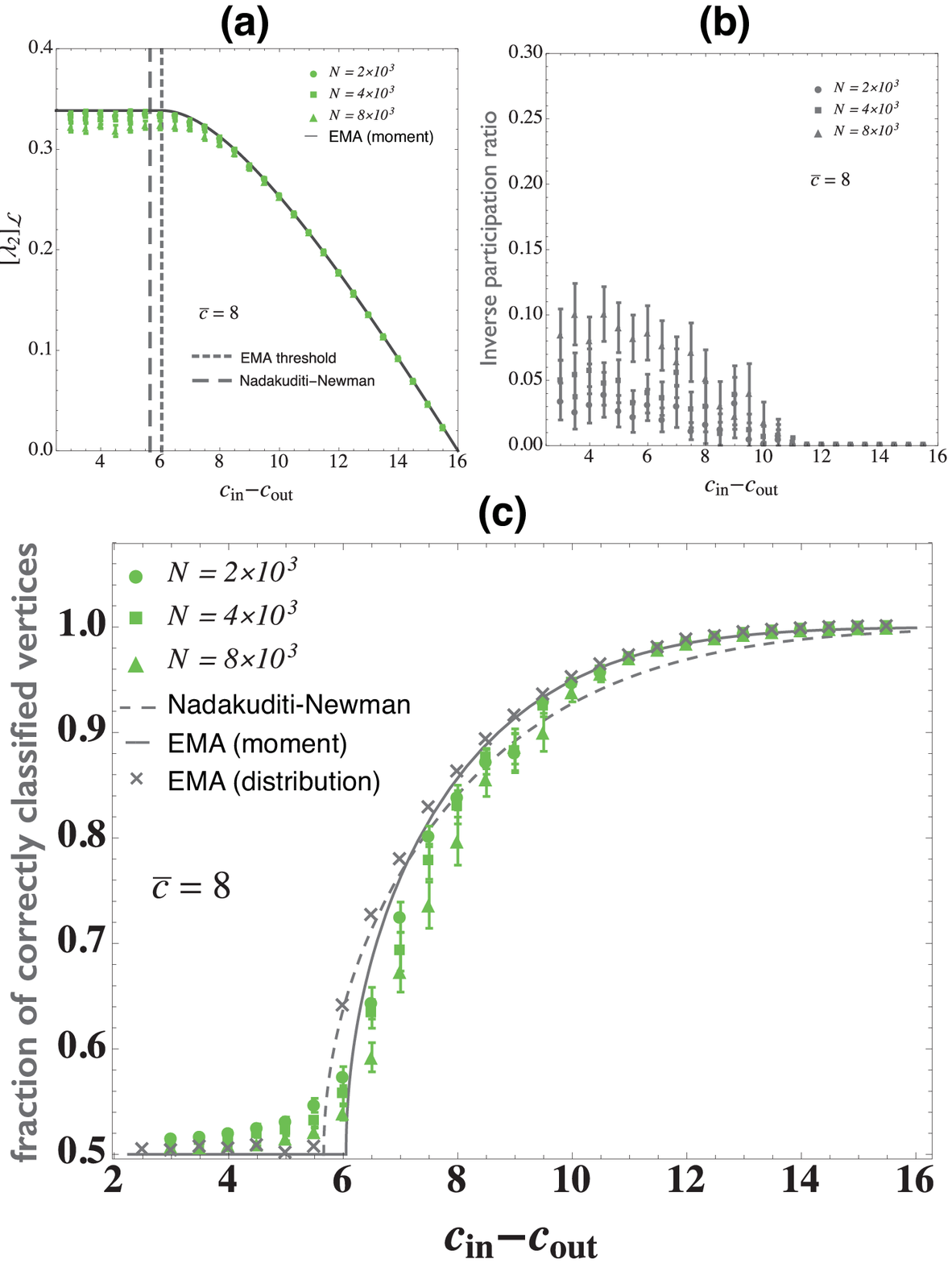}
\caption{
(Color online)
Same plots as in Fig.~\ref{SBM-cm6}, but with the average degree $\overline{c} = 8$.
In this case, the effect of localization is weaker, and the estimate with the EMA is more precise.
}
\label{SBM-cm8}
\end{figure}

We can compare the  average estimate of the second-smallest eigenvalue $\left[\lambda_{2}\right]_{\mathcal{L}}$ with the numerical results.
In terms of $c_{\mathrm{in}}$ and $c_{\mathrm{out}}$, Eq.~(\ref{Ncut-distributed-2ndEigenValue-EMA-2}) reads
\begin{align}
\left[\lambda_{2}\right]_{\mathcal{L}}
&= 1 - \frac{(\overline{c}-1)}{2\overline{c}^{2}} \left( c_{\mathrm{in}} - c_{\mathrm{out}} \right) - \frac{2}{c_{\mathrm{in}} - c_{\mathrm{out}}}. 
\end{align}
Figures \ref{SBM-cm6}(a) and \ref{SBM-cm8}(a) show that the estimated eigenvalues with the EMA agree excellently with the numerical results, as long as the localized eigenstate does not occupy the second-smallest eigenvalue.
We measured the localization strength with the \textit{inverse participation ratio} (IPR), defined as $\sum_{i=1}^{N} x_{i}^{4}/(\sum_{i=1}^{N} x_{i}^{2})^{2}$ for a vector $\bm{x}$, and have plotted this in Figs.~\ref{SBM-cm6}(b) and \ref{SBM-cm8}(b).
The IPR grows rapidly below the point at which the estimates start to deviate from the numerical results in Figs.~\ref{SBM-cm6}(a) and \ref{SBM-cm8}(a).
Similarly, as shown in Figs.~\ref{SBM-cm6}(c) and \ref{SBM-cm8}(c), our estimates for the fraction of correctly classified vertices start to deviate from the  numerical results at that point.

For the stochastic block model with $\overline{c} = 6$, the localized eigenvector appears in the region significantly above the detectability threshold, i.e., in the detectable region, and its effect is not negligible.
However, for the stochastic block model with $\overline{c} = 8$, although the estimate with the EMA is still not precise, the error due to  localization seems to be much smaller for the graph sizes we tested.
Note that as the average degree $\overline{c}$ increases the degree fluctuation of each vertex decreases because of the law of large numbers. Therefore, the effect of localization is expected to  eventually disappear.
That is, the point at which localization occurs will finally become buried in the undetectable region.

As mentioned in Sec.~\ref{DetectabilityNcut}, the solid lines and crosses behave differently in Figs.~\ref{SBM-cm6}(c) and \ref{SBM-cm8}(c).
Moreover, the results of the saddle-point equation with the EMA seem to converge to the result in \cite{Nadakuditi2012} around the detectability threshold.
It should be noted, however, that this coincidence is due to the property of the Poisson degree distribution in the stochastic block model.
As  described in Appendix~\ref{EMAdifference}, the difference between these results with the EMA generally depends on the form of the degree distribution.

%
%
%

%
%
%

\section{Summary} \label{Summary}
In summary, we have analyzed the limitations of the spectral method for graph partitioning, known as the detectability threshold, and the localization of eigenvectors.
We derived estimates for the detectability thresholds of the spectral method with the unnormalized Laplacian $L$ [Eq.~(\ref{transitionpoint-RatioCutEMA-formal})] and the normalized Laplacian $\mathcal{L}$ [Eq.~(\ref{NcutDetectability1})] for sparse graphs.
The detectability threshold with the normalized Laplacian $\mathcal{L}$ can generally be written as (\ref{NcutDetectability2}) (for  equal size modules), which is analogous to the threshold for random regular graphs (\ref{Detectability-3}). This converges to the result in \cite{Nadakuditi2012} in the dense limit $\overline{c} \rightarrow \infty$.
For the  condition where a localized eigenvector emerges, although it is difficult for a graph with an arbitrary degree distribution, our estimates give a fairly good prediction for two-block random graphs with bimodal degree distributions.
Overall, our estimates with the replica method agree with the  numerical results quite well, as long as localization is absent, for the graph sizes we tested.
It should be noted, however, that the localization of  eigenvectors is expected to be sensitive to rare events such as the emergence of  vertices of irregular degree. Therefore, when the support of the degree distribution is infinite, e.g., a Poisson distribution, we must be careful as the finite size effect may not be negligible.

We revealed that the spectral method with Laplacians does not detect modules all the way down to the ultimate detectability threshold $2 \sqrt{\overline{c}}$ in any sparse graph.
In fact, the estimated gap between the precise detectability threshold and the ultimate one is already considerable for very sparse graphs even in the case where eigenvector localization  is absent or negligible (see Fig.~\ref{ThresholdComparison}).
Another way of viewing this result is that the method with the non-backtracking matrix closed the gap of detectability in two ways. 
Furthermore, for the graph sizes we tested, the effect of localization was relatively weak in the stochastic block model when the average degree was not very low.
Finally, we comment that we must be careful when we compare the performance between the normalized Laplacian and modularity. Although their spectral methods become equivalent for a certain choice of normalization \cite{Newman2013}, there is no guarantee that our results precisely coincide with the detectability threshold of the modularity matrix, i.e., the method with the spherical normalization \cite{Newman2006}.

\section*{Acknowledgements}
This work was supported by
JSPS KAKENHI No. 26011023 (TK) and No. 25120013 (YK) and
the JSPS Core-to-Core Program ``Non-equilibrium dynamics of soft matter and information.''

\appendix

\section{Number of random graphs with two modules} \label{GraphCountSingleDegree}
In this section, we calculate $\mathcal{N}_{G}$, the number of possible graph realizations.
Although the value of $\mathcal{N}_{G}$ does not appear in the final result in the main text, the technique used here is essential for calculating the moment of the partition function.
Let us first consider the two-block $c$-random regular graphs.
We sum all the connection patterns $(\{ u_{ij} \}, \{ w_{ij} \})$ that satisfy the constraint of constant degree.
That is,
\begin{align}
\mathcal{N}_{G} = \sum_{ \{u_{ij}\} \{w_{ij}\} }
\prod_{i \in V_{1}} \delta\left( \sum_{l \in V_{1}} u_{il} + \sum_{k \in V_{2}} w_{ik} - c \right)
\prod_{j \in V_{2}} \delta\left( \sum_{l \in V_{2}} u_{jl} + \sum_{k \in V_{1}} w_{jk} - c \right)
\delta\left( \sum_{i \in V_{1}} \sum_{k \in V_{2}} w_{ik} - \gamma N \right).
\end{align}
Using the relations
\begin{align}
&\delta(x) = \oint \frac{dz}{2\pi} \, z^{x-1}, \\
&\delta(x) = \int_{-i\infty}^{+i\infty} \frac{d\eta}{2\pi} \, \mathrm{e}^{-\eta x},
\end{align}
we have
\begin{align}
\mathcal{N}_{G}
&= \sum_{ \{u_{ij}\} \{w_{ij}\} }
\prod_{i \in V_{1}} \oint \frac{dz_{i}}{2\pi} \, z_{i}^{(\sum_{l \in V_{1}} u_{il} + \sum_{k \in V_{2}} w_{ik} - c - 1)}
\prod_{j \in V_{2}} \oint \frac{dz_{j}}{2\pi} \, z_{j}^{(\sum_{l \in V_{2}} u_{jl} + \sum_{k \in V_{1}} w_{jk} - c - 1)} \nonumber\\
&\hspace{20pt} \int \frac{d\eta}{2\pi} \, \exp \bigl( -\eta(\sum_{i \in V_{1}} \sum_{k \in V_{2}} w_{ik} - \gamma N) \bigr) \label{NG1}\\
&=
\oint \prod_{i \in V_{1}} \frac{dz_{i}}{2\pi} z_{i}^{-(1+c)}
\oint \prod_{j \in V_{2}} \frac{dz_{j}}{2\pi} z_{j}^{-(1+c)}
\int \frac{d\eta}{2\pi} \mathrm{e}^{\eta \gamma N} \nonumber\\
&\hspace{20pt} \times
\prod_{i<l \in V_{1}} \sum_{u_{il} = \{0,1\}} (z_{i}z_{l})^{u_{il}}
\prod_{j<l \in V_{2}} \sum_{u_{jl} = \{0,1\}} (z_{j}z_{l})^{u_{jl}}
\prod_{i \in V_{1}} \prod_{k \in V_{2}} \sum_{w_{ik} = \{0,1\}} (z_{i}z_{k}\mathrm{e}^{-\eta})^{w_{ik}} \label{NG4}\\
&=
\oint \prod_{i \in V_{1}} \frac{dz_{i}}{2\pi} z_{i}^{-(1+c)}
\oint \prod_{j \in V_{2}} \frac{dz_{j}}{2\pi} z_{j}^{-(1+c)}
\int \frac{d\eta}{2\pi} \mathrm{e}^{\eta \gamma N} \nonumber\\
&\hspace{20pt} \times
\prod_{i<l \in V_{1}} \left(1 + z_{i}z_{l}\right)
\prod_{j<l \in V_{2}} \left(1 + z_{j}z_{l}\right)
\prod_{i \in V_{1}} \prod_{k \in V_{2}} \left(1 + z_{i}z_{k}\mathrm{e}^{-\eta}\right) \label{NG5}.
\end{align}
Setting the contours with respect to $z_{i}$ and $z_{j}$ to be small, we can approximate the last factors as
\begin{align}
&\prod_{i<l \in V_{1}} (1 + z_{i}z_{l})
= \exp \left( \sum_{i<l \in V_{1}} \ln (1 + z_{i}z_{l}) \right)
\approx \exp \left( \sum_{i<l \in V_{1}} z_{i}z_{l} \right)
\approx \exp \left( \frac{1}{2} \sum_{i \in V_{1}} z_{i} \sum_{l \in V_{1}} z_{l} \right), \label{exponentiate1}\\
&\prod_{i \in V_{1}} \prod_{k \in V_{2}} \left(1 + z_{i}z_{k}\mathrm{e}^{-\eta}\right)
= \exp \left( \sum_{i \in V_{1}} \sum_{k \in V_{2}} \ln (1 + z_{i}z_{k}\mathrm{e}^{-\eta}) \right)
\approx \exp \left( \mathrm{e}^{-\eta} \sum_{i \in V_{1}} z_{i} \sum_{k \in V_{2}} z_{k} \right) \label{exponentiate2},
\end{align}
where we have neglected the diagonal term in (\ref{exponentiate1}) because $N \gg 1$.
Introducing the order parameters
\begin{align}
&q_{1} = \frac{1}{p_{1}N} \sum_{i \in V_{1}} z_{i}, \label{Appendix-OrderParameter1}\\
&q_{2} = \frac{1}{p_{2}N} \sum_{j \in V_{2}} z_{j}, \label{Appendix-OrderParameter2}
\end{align}
we can recast (\ref{NG5}) as
\begin{align}
\mathcal{N}_{G}
&=
p_{1}p_{2} N^{2} \int dq_{1} \int dq_{2}
\oint \prod_{i \in V_{1}} \frac{dz_{i}}{2\pi} z_{i}^{-(1+c)}
\oint \prod_{j \in V_{2}} \frac{dz_{j}}{2\pi} z_{j}^{-(1+c)}
\int \frac{d\eta}{2\pi} \mathrm{e}^{\eta \gamma N} \nonumber\\
&\hspace{20pt} \times \delta \left( p_{1}N q_{1} - \sum_{i \in V_{1}} z_{i} \right)
\delta \left( p_{2}N q_{2} - \sum_{j \in V_{2}} z_{j} \right) \nonumber\\
&\hspace{20pt} \times
\exp\left[ \frac{1}{2} \left( p_{1}N q_{1}\right)^{2} \right]
\exp\left[ \frac{1}{2} \left( p_{2}N q_{2}\right)^{2} \right]
\exp \left[ \mathrm{e}^{-\eta} p_{1}p_{2}N^{2} q_{1}q_{2} \right] \label{NG6} \\
&=
p_{1}p_{2} N^{2} \int \frac{dq_{1} d\hat{q}_{1}}{2\pi} \int \frac{dq_{2} d\hat{q}_{2}}{2\pi}
\oint \prod_{i \in V_{1}} \frac{dz_{i}}{2\pi} z_{i}^{-(1+c)}
\oint \prod_{j \in V_{2}} \frac{dz_{j}}{2\pi} z_{j}^{-(1+c)}
\int \frac{d\eta}{2\pi} \nonumber\\
&\hspace{20pt} \times
\exp \left[ -\hat{q}_{1} \left( p_{1}N q_{1} - \sum_{i \in V_{1}} z_{i} \right) \right]
\exp \left[ -\hat{q}_{2} \left( p_{2}N q_{2} - \sum_{j \in V_{2}} z_{j} \right) \right] \nonumber\\
&\hspace{20pt} \times
\exp\left[ \frac{1}{2} \left( p_{1}N q_{1}\right)^{2}
+ \frac{1}{2} \left( p_{2}N q_{2}\right)^{2}
+ \mathrm{e}^{-\eta} p_{1}p_{2}N^{2} q_{1}q_{2} + \eta \gamma N \right]. \label{NG7}
\end{align}
Since
\begin{align}
\oint \prod_{i \in V_{r}} \frac{dz_{i}}{2\pi} \mathrm{e}^{z_{i}\hat{q}_{r}} z_{i}^{-(1+c)}
&= \oint \prod_{i \in V_{r}} \frac{dz_{i}}{2\pi} \sum_{m} \frac{\left( z_{i}\hat{q}_{r} \right)^{m}}{m!} z_{i}^{-(1+c)}
= \left( \frac{\hat{q}^{c}_{r}}{c !} \right)^{Np_{r}} \hspace{20pt} (r=1,2),
\end{align}
Eq.~(\ref{NG7}) becomes
\begin{align}
\mathcal{N}_{G}
&=
p_{1}p_{2} N^{2} \int \frac{dq_{1} d\hat{q}_{1}}{2\pi} \int \frac{dq_{2} d\hat{q}_{2}}{2\pi}
\int \frac{d\eta}{2\pi}
\exp\Biggl[
\frac{N^{2}}{2} \left( p^{2}_{1} q^{2}_{1} + p^{2}_{2} q^{2}_{2} + 2\mathrm{e}^{-\eta} p_{1}p_{2} q_{1}q_{2} \right) \nonumber\\
&\hspace{100pt} + N \left( \eta \gamma - p_{1} \hat{q}_{1} q_{1} - p_{2} \hat{q}_{2} q_{2} + c \, p_{1} \ln \hat{q}_{1} + c \, p_{2} \ln \hat{q}_{2} - \ln c ! \right)
\Biggr]. \label{NG8}
\end{align}
In the limit $N\rightarrow\infty$, the saddle point of the integrand gives $\mathcal{N}_{G}$.
The saddle-point conditions yield
\begin{align}
& \gamma = N \mathrm{e}^{-\eta} p_{1}p_{2} q_{1}q_{2}, \label{etagraph} \\
& N p_{1} q_{1} + N p_{2}\mathrm{e}^{-\eta}q_{2} - \hat{q}_{1} = 0, \\
& N p_{2} q_{2} + N p_{1}\mathrm{e}^{-\eta}q_{1} - \hat{q}_{2} = 0, \\
& q_{1} \hat{q}_{1} = q_{2} \hat{q}_{2} = c.
\end{align}
We then have
\begin{align}
q_{1} = \sqrt{\frac{c \, p_{1} - \gamma}{N p^{2}_{1}}}, \label{T1} \\
q_{2} = \sqrt{\frac{c \, p_{2} - \gamma}{N p^{2}_{2}}}. \label{T2}
\end{align}
Inserting the values at the saddle point, we obtain the number of graphs for $N \gg 1$:
\begin{align}
\mathcal{N}_{G}
&\simeq \exp \Biggl[
N \biggl(
\frac{\overline{c}}{2} \left( \ln N - 1 \right) - \ln c!
- \gamma \ln \gamma
+ c p_{1} \ln \left( c p_{1} \right)
+ c p_{2} \ln \left( c p_{2} \right) \nonumber\\
& \hspace{50pt} -\frac{1}{2} \left( c p_{1} - \gamma \right) \ln \left( c p_{1} - \gamma \right)
-\frac{1}{2} \left( c p_{2} - \gamma \right) \ln \left( c p_{2} - \gamma \right)
\biggr) \Biggr].
\end{align}

A completely analogous result holds for the number of possible two-block random graphs with a given degree sequence, i.e., the two-block configuration model or the degree-corrected ensembles with ``hard'' constraints \cite{Peixoto2012}.
We let $\{ c_{t} \}_{t=1}^{T}$ be the sequence of degrees, each of which has probability $\{ b_{t} \}_{t=1}^{T}$; i.e., the number of nodes with degree $c_{t}$ is $N b_{t}$.
The number of graphs $\mathcal{N}_{G}$ can then be written as
\begin{align}
\mathcal{N}_{G} = \sum_{ \{u_{ij}\} \{w_{ij}\} } \prod_{t=1}^{T} \left[
\prod_{i \in V_{(1,t)}} \delta\left( \sum_{l \in V_{1}} u_{il} + \sum_{k \in V_{2}} w_{ik} - c_{t} \right)
\prod_{j \in V_{(2,t)}} \delta\left( \sum_{l \in V_{2}} u_{jl} + \sum_{k \in V_{1}} w_{jk} - c_{t} \right) \right]
\delta\left( \sum_{i \in V_{1}} \sum_{k \in V_{2}} w_{ik} - \gamma N \right),
\end{align}
where, as in the main text, we denote  the set of vertices in module $r$ with degree $c_{t}$ as $V_{(r, t)}$.
A similar calculation to the single-degree case yields
\begin{align}
\mathcal{N}_{G} &=
p_{1}p_{2} N^{2} \int \frac{d\eta}{2\pi} \int \frac{dq_{1} d\hat{q}_{1}}{2\pi} \int \frac{dq_{2} d\hat{q}_{2}}{2\pi}
\prod_{t} \left[ \prod_{i \in V_{(1,t)}} \frac{1}{c_{t}!} \prod_{j \in V_{(2,t)}} \frac{1}{c_{t}!} \right] \nonumber\\
&\hspace{50pt}\times\exp\Biggl[
N \biggl( \frac{N}{2} p^{2}_{1} q^{2}_{1} + \frac{N}{2} p^{2}_{2} q^{2}_{2} + N \mathrm{e}^{-\eta} p_{1}p_{2} q_{1}q_{2} \nonumber\\
&\hspace{100pt} + \eta \gamma - p_{1} \hat{q}_{1} q_{1} - p_{2} \hat{q}_{2} q_{2}
+ \overline{c} \left( p_{1} \ln \hat{q}_{1} + p_{2} \ln \hat{q}_{2} \right) \biggr)
\Biggr],
\end{align}
where, as in the main text, $\overline{c} = \sum_{b_{t}} b_{t} c_{t}$ is the average degree.
The saddle-point conditions yield analogous results, 
\begin{align}
& \gamma = N p_{1}q_{1} p_{2}q_{2} \mathrm{e}^{-\eta}, \label{distributed-etagraph}\\
& N p_{1} q_{1} + N p_{2} q_{2} \mathrm{e}^{-\eta} = \hat{q}_{1}, \\
& N p_{1} q_{1} \mathrm{e}^{-\eta} + N p_{2} q_{2} = \hat{q}_{2}, \\
& q_{1} \hat{q}_{1} = q_{2} \hat{q}_{2} = \overline{c},
\end{align}
and
\begin{align}
q_{1} = \sqrt{\frac{\overline{c} p_{1} - \gamma}{N p^{2}_{1}}}, \label{distributedT1} \\
q_{2} = \sqrt{\frac{\overline{c} p_{2} - \gamma}{N p^{2}_{2}}}. \label{distributedT2}
\end{align}
Finally, we have
\begin{align}
\mathcal{N}_{G}
&\simeq \exp \Biggl[
N \biggl(
\frac{\overline{c}}{2} \left( \ln N - 1 \right) - \ln \overline{c!}
- \gamma \ln \gamma
+ \overline{c} p_{1} \ln \left( \overline{c} p_{1} \right)
+ \overline{c} p_{2} \ln \left( \overline{c} p_{2} \right) \nonumber\\
& \hspace{50pt} -\frac{1}{2} \left( \overline{c} p_{1} - \gamma \right) \ln \left( \overline{c} p_{1} - \gamma \right)
-\frac{1}{2} \left( \overline{c} p_{2} - \gamma \right) \ln \left( \overline{c} p_{2} - \gamma \right)
\biggr) \Biggr],
\end{align}
where $\overline{c!} = \sum_{t} b_{t} c_{t}!$.

\section{Gaussian approximation of the fraction of correctly classified vertices} \label{Appendix-CorrectFraction}
We consider the distribution of elements of the second-smallest eigenvector belonging to the $r$th module, averaged over the realization of the unnormalized Laplacian $L$.
This is defined as
\begin{align}
P_{r}(x) = \frac{1}{N_{r}} \sum_{i \in V_{r}} \left[ \delta(x - x_{i}) \right]_{L}, \label{Appendix-eigendistribution1}
\end{align}
and, as  mentioned in the main text, it can be expressed in terms of $H$ and $A$ as
\begin{align}
P_{r}(x) = \int dA dH \, Q_{r}(A,H) \, \delta\left(x - \frac{H}{A}\right). \label{Appendix-eigendistribution2}
\end{align}
For the Gaussian fitting of the distribution $P_{r}(x)$, we solve for the mean and variance in (\ref{Appendix-eigendistribution2}).
To obtain an analytical expression, we fix the distribution of $A$, i.e., $Q_{r}(A) = \int dH Q_{r}(A,H) = \delta(a_{\mathrm{full}} - A)$.
Note that $a_{\mathrm{full}}$ here is different from $a$ in the saddle-point equations.
From the cavity interpretation of Eqs.~(\ref{cavity1-Regular})--(\ref{cavity3-Regular}), $Q_{r}(A,H)$ can be regarded as the complete marginal distribution corresponding to $q_{r}(A,H)$.
Therefore, instead of (\ref{saddlepoint-a-RegularRandom}), $a_{\mathrm{full}}$ can be determined as
\begin{align}
a_{\mathrm{full}} &= \phi - c \hat{a} \nonumber\\
&= (c-1) \Gamma - \frac{1}{(c-1) \Gamma}, \label{Appendix-afull}
\end{align}
where we have inserted the values of $\phi$ and $\hat{a}$ at the saddle point.
Then, in the detectable region, the mean can be approximated as
\begin{align}
m_{r}(x) &= \frac{1}{a_{\mathrm{full}}} \int dH \, H Q_{r}(H) \nonumber\\
&= \frac{1}{a_{\mathrm{full}}} \int \prod_{g=1}^{c} d\hat{H}_{g} \hat{q}_{r}(\hat{H}_{g}) \left( -\frac{\psi}{2} + \sum_{g=1}^{c} \hat{H}_{g} \right) \nonumber\\
&= \frac{ c (c-1) \Gamma \hat{m}_{1r} }{(c-1)^{2} \Gamma^{2} - 1},
\end{align}
where we have used the fact that $\psi = 0$ at the saddle point.

Similarly, for the second moment, we have
\begin{align}
\bracket{x^{2}}_{r} &= \frac{1}{a^{2}_{\mathrm{full}}} \int \prod_{g=1}^{c} d\hat{H}_{g} \hat{q}_{r}(\hat{H}_{g}) \left( \sum_{g=1}^{c} \hat{H}_{g} \right)^{2} \nonumber\\
&= \left(\frac{ (c-1) \Gamma }{(c-1)^{2} \Gamma^{2} - 1}\right)^{2} \left( c \hat{m}_{2r} + c(c-1) \hat{m}^{2}_{1r} \right). \label{GaussianMean-Regular}
\end{align}
Then, the variance reads
\begin{align}
s^{2}_{r}(x) &= \bracket{x^{2}}_{r} - \bracket{x^{2}}_{r} \nonumber\\
&= c \left(\frac{ (c-1) \Gamma }{(c-1)^{2} \Gamma^{2} - 1}\right)^{2} \left( \hat{m}_{2r} - \hat{m}^{2}_{1r} \right). \label{GaussianVariance-Regular}
\end{align}
With $m_{r}(x)$ and $s_{r}(x)$, $\left[ 1 + \mathrm{erf}(m_{r}(x)/\sqrt{2 s^{2}_{r}(x)}) \right]/2$ gives the fraction of correctly classified vertices in module $r$.
For the equal-size modules, $|m_{1}(x)|=|m_{2}(x)| (=:|m(x)|)$ and $s^{2}_{1}(x)=s^{2}_{2}(x) (=:s^{2}(x))$ by symmetry.
The total fraction of correctly classified vertices is then $\left[ 1 + \mathrm{erf}(|m(x)|/\sqrt{2 s^{2}(x)}) \right]/2$.

For the normalized Laplacian $\mathcal{L}$ with the EMA, the analogous calculation for the detectable region gives
\begin{align}
m_{r}(x) &= \frac{\overline{c^{2}}}{\overline{c}} \frac{  (\overline{c}-1) \overline{\Gamma} \hat{m}_{1r} }{(\overline{c}-1)^{2} \overline{\Gamma}^{2} - 1}, \label{GaussianMean-NcutEMA}\\
s^{2}_{r}(x) &= \left( \frac{  (\overline{c}-1) \overline{\Gamma} }{(\overline{c}-1)^{2} \overline{\Gamma}^{2} - 1} \right)^{2}
\left[ \frac{\overline{c^{2}}}{\overline{c}} \hat{m}_{2r} + \left( \frac{\overline{c^{3}} - \overline{c^{2}}}{\overline{c}} - \left( \frac{\overline{c^{2}}}{\overline{c}} \right)^{2} \right) \hat{m}^{2}_{1r} \right], \label{GaussianVariance-NcutEMA}
\end{align}
where $\overline{c^{n}} = \sum_{t} b_{t} c_{t}^{n}$.

\section{Approximations of the saddle-point equations and the free energy}\label{EMAdifference}
As shown in Figs.~\ref{NcutCorrectFractionBimodal}(a), \ref{NcutCorrectFractionBimodal}(b), \ref{SBM-cm6}(c), and \ref{SBM-cm8}(c), the results obtained by the EMA of the saddle-point equation (crosses in the figures) and the Gaussian fitting of the distribution of the eigenvector elements (solid lines in the figures), whose mean and variance are estimated by the EMA of the free energy, are different.
One may expect that the former is simply more accurate, as long as the stationary state is achieved by a sufficient number of iterations, because the latter contains a Gaussian approximation.
However, this is not correct.
In fact, the EMA of the saddle-point equations is not equivalent to the EMA of the free energy; in the latter, the approximation is applied before taking the saddle point.
For example, the relation between the first moment of $H$, $m_{1r}$, and its conjugate $\hat{m}_{1r}$, derived by the saddle-point equation (\ref{cavity3-Ncut}) with the EMA, is
\begin{align}
m_{1r} = \left( \frac{\overline{c^{2}}}{\overline{c}} - 1 \right) \hat{m}_{1r}, \label{SaddlePointEMA}
\end{align}
whereas we have $m_{1r} = (\overline{c}-1) \hat{m}_{1r}$ from the saddle point of the free energy with the EMA.
Hence, the difference depends on the ratio of the mean and the variance, or the Fano factor.
In the case of the Poisson degree distribution, Eq.~(\ref{SaddlePointEMA}) reads $m_{1r} = \overline{c} \hat{m}_{1r}$, which corresponds to the dense limit of the saddle point for the free energy.
As neither quantity is exact, it is not readily obvious which offers the better estimate in general.

\bibliographystyle{apsrev}

\bibliography{DetectabilityLocalization}

\begin{thebibliography}{46}
\bibitem[{Com()}]{CommentNcutBimodal}
\bibinfo{howpublished}{In Fig.~8(a), the localization transition might have
  occurred right before the detectability threshold. However, we could not find
  an eigenvalue close to the numerial experiment within $g \le 3$.}
\end{thebibliography}



\end{document}